\documentclass[11pt,a4paper]{article}
\usepackage{jcappub}

\usepackage{url}
\usepackage{lscape}

\def\beq{\begin{equation}}
\def\eeq{\end{equation}}
\def\alwaysmath#1{{\ifmmode{#1}\else{$#1$}\fi}}
\def\he#1{\hbox{\alwaysmath{{}^{#1}}{\rm He}}}

\def\etal{{\it et al.}~}
\def\sun{${\,_\odot}$}

\title{
An MCMC determination of the primordial helium abundance
}

\author[a]{Erik Aver}
\author[a,b,c]{Keith~A.~Olive}
\author[a,c]{Evan~D.~Skillman}

\affiliation[a]{School of Physics and Astronomy, University of Minnesota, \\
116 Church St. SE, Minneapolis, MN 55455}
\emailAdd{aver@physics.umn.edu}
\affiliation[b]{William I. Fine Theoretical Physics Institute, University of Minnesota, \\
116 Church St. SE, Minneapolis, MN 55455}
\emailAdd{olive@umn.edu}
\affiliation[c]{Minnesota Institute for Astrophysics, University of Minnesota, \\
116 Church St. SE, Minneapolis, MN 55455}
\emailAdd{skillman@astro.umn.edu}

\abstract{
Spectroscopic observations of the chemical abundances in metal-poor H~II regions provide an
independent method for estimating the primordial helium abundance.
H~II regions are described by several physical parameters such as electron density, 
electron temperature, and reddening, in addition to y, the ratio of helium to hydrogen.  
It had been customary to estimate or determine self-consistently these parameters to calculate y.  
Frequentist analyses of the parameter space have been shown to be successful in these parameter determinations, and Markov Chain Monte Carlo (MCMC) techniques have proven to be very efficient in sampling this parameter space.  Nevertheless, accurate determination of the primordial helium abundance from observations of H~II regions is constrained by both systematic and statistical uncertainties. 
In an attempt to better reduce the latter, and continue to better characterize the former,  
we apply MCMC methods to the large dataset recently compiled by Izotov, Thuan, \& Stasi\'nska (2007). 
To improve the reliability of the determination, a high quality dataset is needed.  In pursuit of this, a variety of cuts are explored.  
The efficacy of the He~I $\lambda$4026 emission line as a constraint on the solutions is first examined, 
revealing the introduction of systematic bias through its absence.  
As a clear measure of the quality of the physical solution, a $\chi^2$ analysis proves instrumental in the selection of data compatible with the theoretical model.  
In addition, the method also allows us to exclude systems for which parameter estimations are
statistical outliers.  As a result, the final selected dataset gains in reliability and exhibits improved consistency.  Regression to zero metallicity yields Y$_{p}$ $=$ 0.2534 $\pm$ 0.0083, in broad agreement with the WMAP result.  The inclusion of more observations shows promise for further reducing the uncertainty, but more high quality spectra are required.
}


\arxivnumber{}

\begin{document}

\begin{flushright}UMN-TH-3022/11\\FTPI-MINN-11/30\\
%astro-ph/0405588 \\
December 2011\end{flushright}
\vskip -0.68in

\maketitle
\flushbottom

\section{Introduction}

Next to the cosmic microwave background radiation, standard big bang nucleosynthesis (SBBN)
is the most robust probe of the early universe available  \citep{wssok,osw,fs}.
Furthermore, using the precise baryon density as determined by WMAP  \citep{wmap,wmap10},
SBBN has effectively become a parameter free theory \cite{cfo2}. As such, one can
use SBBN to make relatively precise predictions of the initial light element abundances of 
 D, $^{3}$He, $^{4}$He, and $^{7}$Li \citep{cfo,coc,cfo3,coc2,cyburt,coc3,cuoco,serp,cfo5,coc4}.
Therefore, an observational determination of these abundances becomes a test
 of the concordance between SBBN theory and the analyses of microwave background
 anisotropies. To test these predictions, the observed abundances must be determined with relatively high precision.  Unfortunately, there is a logarithmic relationship between the baryon to photon ratio, $\eta$, and the primordial helium abundance, Y$_{p}$. Thus, any meaningful test of the theory requires a determination of Y$_{p}$ to an accuracy of $\lesssim 1\%$. The 7-year WMAP value for $\eta$ is $(6.19 \pm 0.15) \times 10^{-10}$ \citep{wmap10}.  For comparison, the SBBN calculation of \citet{cfo5}, assuming the WMAP $\eta$ and a neutron mean life of $885.7 \pm 0.8$ s \citep{rpp}, yields $Y_p = 0.2487 \pm 0.0002$, a relative uncertainty of only 0.08\%.

Here, we discuss the determination of Y$_{p}$ using observations of low metallicity H~II regions in dwarf galaxies.  By fitting the helium abundance versus metallicity, one can extrapolate back to very low metallicity, corresponding to the primordial helium abundance \citep{ptp74}.  The oxygen to hydrogen ratio, O/H, commonly serves as a proxy for metallicity.  The difficulties in calculating an accurate and precise measure of the primordial helium abundance are well established \citep{os01,os04,its07}. Previously, we introduced a new method based on Markov Chain Monte Carlo (MCMC) techniques \citep[][AOS2]{AOS2}, and here, we systematically apply this technique to the data compiled in \citep[][ITS07]{its07}.

Observations of the helium to hydrogen emission line ratios from extragalactic H~II regions provide a measure of the helium to hydrogen ratio, y$^{+}={n(He~II) \over n(H~II)}$.  The uncertainty in y$^{+}$ follows from the statistical measurement errors of the helium and hydrogen emission line fluxes, in addition to a myriad of systematic effects.  Interstellar reddening, underlying stellar absorption, radiative transfer, and collisional corrections alter the observed flux, complicating the determination of y$^{+}$, and amplifying the uncertainty.  First, dust along the line of sight scatters the emitted photons (interstellar reddening).  Second, the stellar continuum juxtaposes absorption features under the emission lines (helium and hydrogen underlying absorption).  Third, before leaving the H~II region the emitted photons can be absorbed and re-emitted (radiative transfer).  Finally, in addition to the dominant recombination emission, collisional excitation also contributes to the emission (collisional corrections for helium and hydrogen).  Because none of these processes is directly measured, they cannot be removed independent of the observed emission lines and theoretical models.  As a result, the uncertainty on  y$^{+}$ must reflect the presence of these systematic effects.  Therefore, high fidelity spectra are required to accurately determine y$^{+}$, while simultaneously estimating the model parameters needed to correct for the listed systematic effects.  This desired confidence weighs against the need for larger sample sizes to decrease the uncertainty on Y$_p$ (and $dY/dZ$).

Motivated by the work of \citet{itl94} and \citet{ppr00}, the importance of Monte Carlo techniques was demonstrated in a ``self-consistent'' analysis method for determining the nebular helium abundance based upon six helium and four hydrogen lines \cite{os01,os04}.   In preceding work, \citet[][AOS]{AOS} updated and extended the physical model and integrated the helium and hydrogen calculations with the goals of increasing accuracy and removing assumptions.  Improving the statistical technique of AOS, \citet[][AOS2]{AOS2} used a Markov Chain Monte Carlo (MCMC) analysis to explore a global likelihood function for all model parameters including the helium abundance.  

The small sample size employed in AOS \& AOS2 implied a clear avenue for improvement.  The small sample had limited capacity to constrain the slope of He vs. O/H and contributed to the large uncertainty on the determination of Y$_p$.  Since the publication of \citet{os04}, Izotov \etal have continued making new observations of H~II regions, significantly increasing their dataset \citep[][]{it04, its07}.  This work makes use of that expanded dataset, aided by the MCMC analysis, which is used in the process of making quality cuts in the dataset.  Included in this process is the use of $\chi^{2}$ as a measure of goodness-of-fit between the model and data.  A new, more rigorously selected dataset holds the promise of an improved determination of the primordial helium abundance.

Section \ref{Model} briefly summarizes the model of AOS \& AOS2.  A broad overview of the ITS07 dataset is given in \S \ref{Sample}, and in \S \ref{4026}, the subsets of the observations with and without the detection of He~I $\lambda$4026 are compared.  Subsequently, further cuts on the dataset are investigated.  First, providing an indication of the quality of the fit, the $\chi^{2}$ of the best-fit solution is scrutinized in \S \ref{Chi}.  Second,  in \S \ref{IllBehaved}, ill-behaved points with unreliable determinations are exhibited and discussed.  Third, \S \ref {Outliers} examines the distribution of the dataset in terms of model parameters used to correct for systematic effects.  After making cuts on the dataset, Y$_p$ is determined in \S \ref{Results}.  Finally, \S \ref{Conclusion} offers a discussion of the results, their exploration, and of further improvements in the determination of the primordial helium abundance.

\section{Model overview} \label{Model}

This work uses the model introduced in AOS and the MCMC statistical analysis introduced in AOS2 and applies them to a larger dataset.  The basic definitions are summarized below.  Please see AOS and AOS2 for full details and discussion.  CosmoMC\footnote{\url{http://cosmologist.info/cosmomc/}} is used to perform the Markov Chain Monte Carlo analysis: efficiently exploring the parameter space and calculating the $\chi^{2}$,
\beq
\chi^2 = \sum_{\lambda} {(\frac{F(\lambda)}{F(H\beta)} - {\frac{F(\lambda)}{F(H\beta)}}_{meas})^2 \over \sigma(\lambda)^2},
\label{eq:X2}
\eeq
where the emission line fluxes, $F(\lambda)$, are measured or calculated for 
six helium lines  ($\lambda$3889, 4026, 4471, 5876, 6678, and 7065) and three
hydrogen lines  ($H\alpha$, $H\gamma$, $H\delta$) each relative to $H\beta$.
The  $\chi^{2}$ in eq. \ref{eq:X2} runs over all He and H lines and $\sigma(\lambda)$ is the
measured uncertainty in the flux ratio at each wavelength.
The best-fit solution (minimum $\chi^{2}$) is then found and frequentist confidence levels are determined from $\Delta\chi^{2}$.

The calculated He flux at each wavelength $\lambda$ relative to the flux in $H\beta$ is given by
\beq
\frac{F(\lambda)}{F(H\beta)}= y^{+}\frac{E(\lambda)}{E(H\beta)}{\frac{W(H\beta)+a_{H}(H\beta)}{W(H\beta)} \over \frac{W(\lambda)+a_{He}(\lambda)}{W(\lambda)}}{f_{\tau}(\lambda)}\frac{1+\frac{C}{R}(\lambda)}{1+\frac{C}{R}(H\beta)}10^{-f(\lambda)C(H\beta)}.
\label{eq:F_He_EW}
\eeq
The ratio of H fluxes is defined analogously, 
\beq
\frac{F(\lambda)}{F(H\beta)}= \frac{E(\lambda)}{E(H\beta)}{\frac{W(H\beta)+a_{H}(H\beta)}{W(H\beta)} \over \frac{W(\lambda)+a_{H}(\lambda)}{W(\lambda)}}\frac{1+\frac{C}{R}(\lambda)}{1+\frac{C}{R}(H\beta)}10^{-f(\lambda)C(H\beta)}.
\label{eq:F_H_EW}
\eeq
The predicted model fluxes shown above are calculated from an input value of y$^{+}$ and an emissivity ratio of the helium or hydrogen line to H$\beta$, $\frac{E(\lambda)}{E(H\beta)}$, with corrections made for reddening (C(H$\beta$)), underlying absorption (a$_{H}$ \& a$_{He}$), collisional enhancement, and radiative transfer.  The optical depth function, \textit{$f_{\tau}$}, and collisional to recombination emission ratio, \textit{$\frac{C}{R}$}, are both temperature (T) and density (n$_{e}$) dependent (the emissivities are also temperature dependent).  The parameters
a$_{He}$ and a$_H$ correspond to $\lambda$4471 and H$\beta$ respectively. The wavelength dependence of the underlying absorption is discussed in detail in AOS.  Additionally, the hydrogen collisional emission depends on the neutral to ionized hydrogen ratio ($\xi$).  Therefore, there are a total of eight model parameters (y$^{+}$, n$_{e}$, a$_{He}$, $\tau$, T, C(H$\beta$), a$_{H}$, $\xi$).  An extensive description and analysis of the physical model is provided in AOS.  The statistical method of sampling the multi-dimensional parameter space is described in AOS2.

The model fluxes also rely on
the measured equivalent widths ($W(\lambda)$). However,
the flux of the continuum at each wavelength, $h(\lambda)$, which relates the line flux to the equivalent width, is constrained such that changes in the equivalent width are proportional to changes in the flux (see AOS2):  
\beq
\frac{h(\lambda)}{h(H\beta)} = {\frac{F(\lambda)}{F(H\beta)}}_{meas} \frac{W(H\beta)_{meas}}{W(\lambda)_{meas}}
\eeq
\beq
\frac{F(\lambda)}{F(H\beta)} = \frac{W(\lambda)}{W(H\beta)}\frac{h(\lambda)}{h(H\beta)}.
\eeq
Finally, the temperature derived from the [O~III] $\lambda\lambda$4363, 4959, and 5007 emission lines, T(O~III), is incorporated as a prior with the resulting term added into the $\chi^{2}$:
\beq
\chi^2_{T}=(T-T(OIII))^2/\sigma^2),
\eeq
where $\sigma=0.2~T_{OIII}$ provides a very weak constraint, but is useful in eliminating unphysical solutions (see AOS2).

\section{The ITS07 sample} \label{Sample}

This work analyzes the HeBCD sample of \citet[][ITS07]{its07}.  The HeBCD sample consists of 93 different observations of 86 H~II regions in 77 galaxies.  As described in ITS07, the spectra were reduced and extracted using IRAF.  The line flux errors are determined from statistical counting uncertainty combined with an assumed standard star absolute flux calibration error of 1\%.   In addition to the helium and hydrogen line flux ratios and equivalent widths, this work makes use of T(O~III) and O/H, as reported in ITS07.  Corrected line fluxes and values for He~I $\lambda$4026 were obtained from Y.~Izotov (private communication).  

\begin{figure}[ht!]
\centering  
\resizebox{\textwidth}{!}{\includegraphics{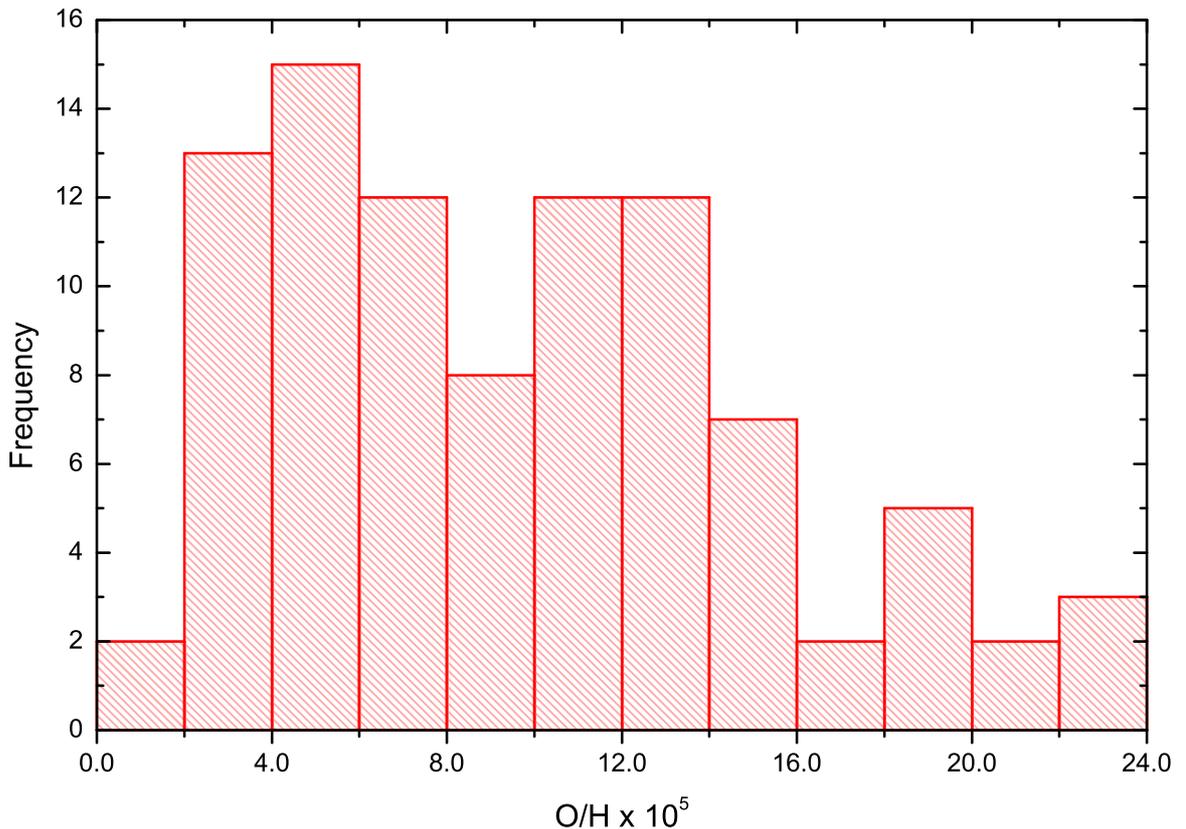}}
\caption{
Histogram of the HeBCD sample of ITS07 in terms of the oxygen to hydrogen ratio, O/H.
}
\label{OH1}
\end{figure}

\begin{figure}[ht!]
\centering  
\resizebox{\textwidth}{!}{\includegraphics{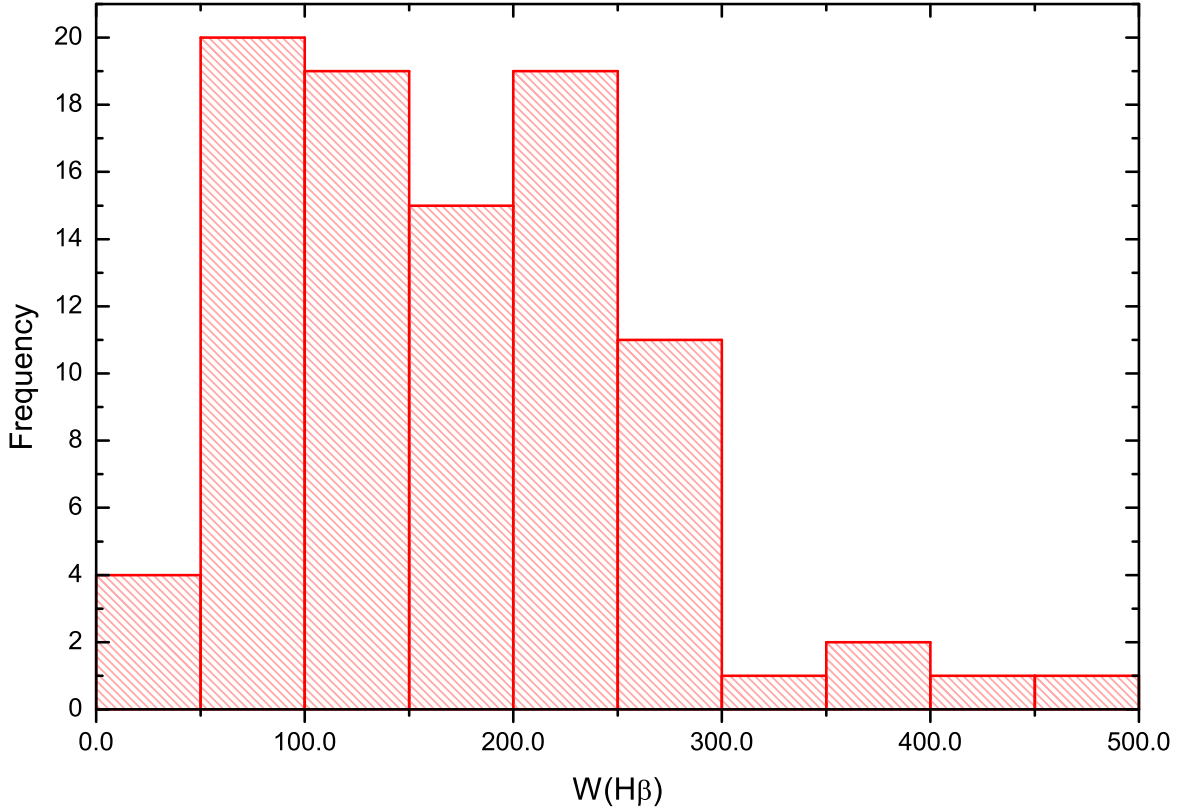}}
\caption{
Histogram of the HeBCD sample of ITS07 in terms of the equivalent width of H$\beta$, W(H$\beta$).
}
\label{EWHB}
\end{figure}

To begin to characterize the sample, the distribution in O/H is shown in figure \ref{OH1}.  As lower metallicities are more desirable for determining Y$_{p}$, the ITS07 sample targeted low metallicities. 
Slightly more than half of the sample have O/H less than the $O/H = 9.2 \times 10^{-5}$ limit imposed in \citet{os04}.  For reference, the solar value is $O/H$\sun$ = (4.90 \pm 0.02) \times 10^{-4}$ \citep{asp09}.  This work does not impose a hard cut-off on O/H.  Instead, first the entire sample is analyzed, then screened for quality, and the distribution of metallicities is examined only as a final step.  Ideally, the density of points along the metallicity baseline should be high enough to capture the uncertainty in the solutions; as a result, any outliers will be excluded.

Figure \ref{EWHB} shows the sample's distribution in W(H$\beta$).  At lower values of W(H$\beta$) the corrections for underlying absorption have a larger effect; as a result, \citet{os04} only used objects with $W(H\beta) > 200$ \AA.  In this work, as with O/H, all objects are analyzed and then investigated for reliability.  In particular, the behavior of objects with large underlying absorption corrections is examined.  Furthermore, given the decreasing signal to noise ratio of low equivalent width objects, in addition to their larger susceptibility to systematic uncertainty introduced by the corrections for underlying absorption, one could imagine identifying a threshold in W(H$\beta$) below which reliable solutions were not possible.  However, we found no such threshold.  Some objects with relatively low equivalent widths ($W(H\beta) \sim 100$) proved to be reliable and showed similar solution characteristics to higher $W(H\beta)$ spectra.

\section{The importance of He~I $\lambda$4026} \label{4026}

For 23 of the 93 observations in the HeBCD sample, the weak line He~I $\lambda$4026 is not detected.  These objects can still be analyzed as described in \S \ref{Model}, with the helium $\chi^2$ composed of the sum of the remaining five He lines.  To help evaluate the reliability of these 23 objects, the impact of removing He~I $\lambda$4026 is investigated in five H~II regions.  These galaxies were chosen to sample a range of equivalent widths:  SBS~0917+527, HS~0924+3821, SBS~1152+579, Mrk~209, and SBS~0940+544~2, with W(H$\beta$) of 85.9, 114.2, 189.3, 224.1, and 241.1 \AA, respectively.  An eight parameter fit to yield the minimum $\chi^{2}$ was performed and uncertainties were calculated from $\Delta\chi^2$.  The results of this analysis are given in table \ref{table:4026}.  Note that without He~I $\lambda$4026, the system is, in principle, over-constrained; though due to degeneracies, a ``perfect $\chi^{2}=0$'' is not found.  The absolute and fractional differences between the results, when solved with and without He~I $\lambda$4026, are also given in table \ref{table:4026}.

From table \ref{table:4026}, we see that
the most significant effect of removing He~I $\lambda$4026 is clearly, and not surprisingly, on the value of a$_{He}$, which increases by 0.23 \AA, on average, 
 which is larger than the typical uncertainty in the solutions including He~I $\lambda$4026.  
The average fractional increase is 69\%.
The uncertainty on a$_{He}$ also increases notably, with the average uncertainty tripling.
From table \ref{table:4026}, it is clear that none of the other determinations of physical conditions show a similar significant bias with the absence of
a He~I $\lambda$4026 measurement\footnote{While a$_H$ also shows a systematic increase when
$\lambda$4026 is removed, the shift is small compared with the original uncertainties.}.  

Since the correction for underlying helium absorption translates linearly into the helium abundance,
as can be seen in eq. \ref{eq:F_He_EW},  
increases in the underlying helium absorption directly lead to an increased y$^{+}$, 
Table \ref{table:4026} shows an increase in y$^{+}$ for all five objects, with an average fractional increase of 4\%. 
Clearly this is an unacceptable bias when the ultimate goal is an uncertainty on the order of $\sim$ 1\%.  

\begin{figure}
\centering  
\resizebox{\textwidth}{!}{\includegraphics{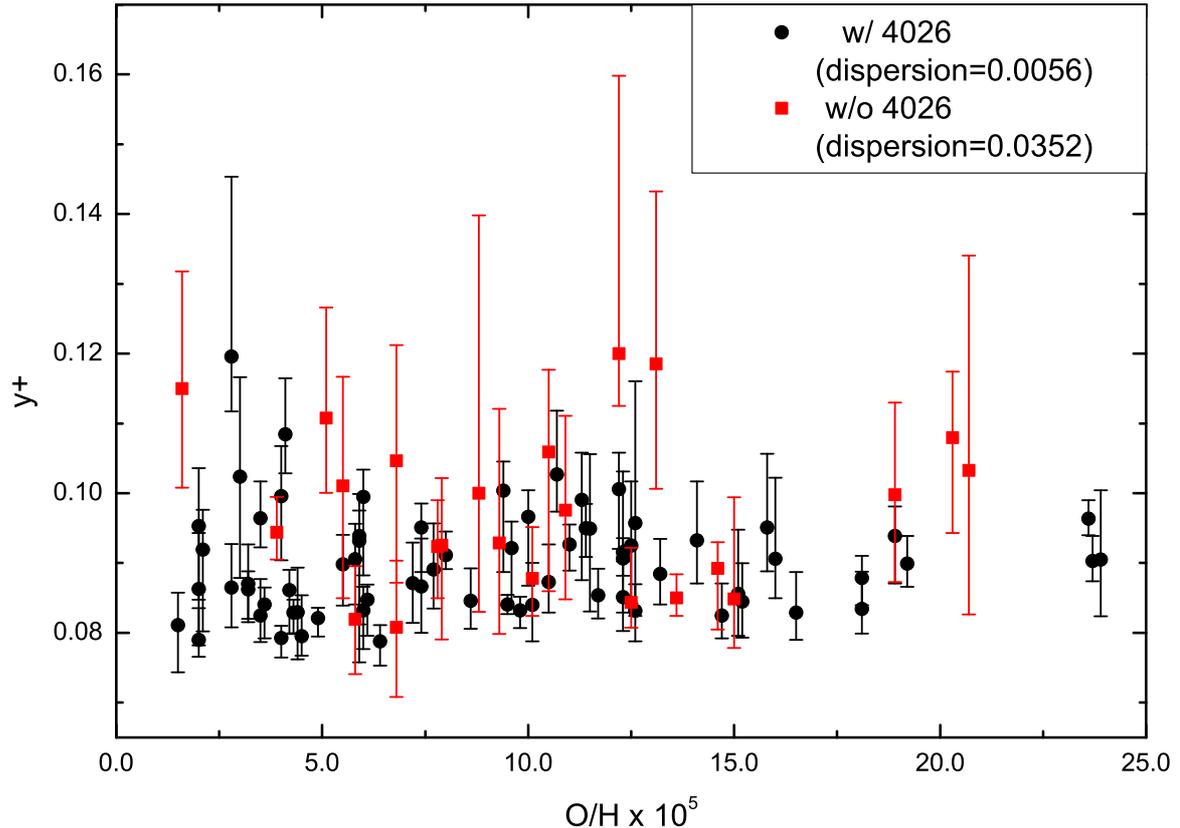}}
\caption{
Plot of the helium abundance versus O/H for the entire sample.  The black circles denote objects for which He~I $\lambda$4026 is reported and was used to calculate $\chi^{2}$ and determine the best-fit solution, while the red squares mark objects for which He~I $\lambda$4026 is not reported and is, therefore, excluded from their analysis.  An upward bias in the value of y$^{+}$ is apparent in the distribution of points lacking He~I $\lambda$4026.
}
\label{y4026}
\end{figure}

This systematic bias is further demonstrated in figure \ref{y4026}, which shows the helium abundances for all 93 objects.  
The set of 23 observations without He~I $\lambda$4026 exhibits bias toward higher values of y$^{+}$, with a subset of these objects noticeably elevated from the majority of the points with He~I $\lambda$4026.  A calculation of the mean for each population underscores the shift; the mean abundance for the 70 objects with He~I $\lambda$4026 is $<y^{+}> = 0.0867 \pm 0.0056$, while it is $<y^{+}> = 0.0900 \pm 0.0352$ for the 23 without He~I $\lambda$4026.  The uncertainties reported with each $<y^{+}>$ are the dispersions of the sample.  The seven-fold increase in dispersion for the sample without He~I $\lambda$4026 demonstrates its unreliability.  

He~I $\lambda$4026, being the weakest of the six helium lines used in our analysis, 
is the most sensitive to underlying helium absorption.
He~I $\lambda$4026 also shows relatively weak sensitivity to temperature, density (collisional 
enhancement), and optical depth, 
and \citet{os01} demonstrated its usefulness for the purpose of constraining
underlying helium absorption.
However, this also
requires that the input spectrum is a very high quality one.  
The addition of even weaker He~I lines such as $\lambda$4922, $\lambda$7281, and $\lambda$4387
\citep[e.g.,][]{ppr00,plp07}, which are all singlet lines with low susceptibility
to optical depth effects, could, in principle, provide even 
stronger constraints on the effects of underlying absorption.  
For this to be true, the spectra would need to be of much higher quality than is typically
obtained for this type of work.

\begin{landscape}
\begin{table}[b!]\footnotesize
%\begin{sidewaystable}[b]\footnotesize
\centering
\vskip .1in
\begin{tabular}{lccccccccc}
%\tabletypesize{\footnotesize}
%\tablewidth{0pt}
\hline
\hline
Object 			& He$^+$/H$^+$     		   &  n$_e$     		   &a$_{He}$     		& $\tau$     				& T$_e$     		    & C(H$\beta$)     	  & a$_{H}$     	       & $\xi$ $\times$ 10$^4$ & $\chi^2$\\
\hline
\hline
\multicolumn{9}{c}{Including He~I $\lambda$4026} \\
\hline
SBS~0917+527    & 0.0866 $^{+0.0069}_{-0.0067}$ &       1 $^{+    196}_{-      1}$ &  0.16 $^{+ 0.13}_{- 0.12}$ &  0.00 $^{+ 0.59}_{- 0.00}$ & 12,770 $^{+1450}_{-1830}$ &  0.05 $^{+ 0.05}_{- 0.05}$ &  0.45 $^{+ 0.65}_{- 0.45}$ &      711 $^{+    5510}_{-     711}$ &  1.2 \\
HS~0924+3821    & 0.0851 $^{+0.0052}_{-0.0049}$ &      92 $^{+    836}_{-     92}$ &  0.35 $^{+ 0.13}_{- 0.12}$ &  0.21 $^{+ 0.82}_{- 0.21}$ & 11,600 $^{+1480}_{-2190}$ &  0.16 $^{+ 0.02}_{- 0.05}$ &  2.29 $^{+ 0.71}_{- 0.52}$ &        0 $^{+   10050}_{-       0}$ &  0.9 \\
SBS~1152+579    & 0.0891 $^{+0.0066}_{-0.0056}$ &       1 $^{+    130}_{-      1}$ &  0.42 $^{+ 0.15}_{- 0.14}$ &  2.48 $^{+ 0.75}_{- 0.61}$ & 15,170 $^{+1840}_{-2040}$ &  0.23 $^{+ 0.04}_{- 0.05}$ &  0.44 $^{+ 0.81}_{- 0.44}$ &       77 $^{+     498}_{-      77}$ &  3.8 \\
Mrk~209         & 0.0848 $^{+0.0022}_{-0.0052}$ &       1 $^{+    293}_{-      1}$ &  0.37 $^{+ 0.11}_{- 0.16}$ &  0.49 $^{+ 0.41}_{- 0.49}$ & 16,030 $^{+1380}_{-2580}$ &  0.01 $^{+ 0.02}_{- 0.01}$ &  2.16 $^{+ 0.95}_{- 0.80}$ &        0 $^{+      59}_{-       0}$ &  1.4 \\
SBS~0940+544~2  & 0.0870 $^{+0.0057}_{-0.0050}$ &       8 $^{+    131}_{-      8}$ &  0.51 $^{+ 0.18}_{- 0.18}$ &  0.08 $^{+ 0.64}_{- 0.08}$ & 18,380 $^{+1330}_{-2500}$ &  0.07 $^{+ 0.03}_{- 0.05}$ &  1.81 $^{+ 1.34}_{- 1.09}$ &        8 $^{+      74}_{-       8}$ &  0.7 \\
\hline
\multicolumn{9}{c}{With He~I $\lambda$4026 Removed} \\
\hline
SBS~0917+527    & 0.0921 $^{+0.0068}_{-0.0115}$ &       1 $^{+    167}_{-      1}$ &  0.34 $^{+ 0.24}_{- 0.34}$ &  0.00 $^{+ 0.54}_{- 0.00}$ & 13,090 $^{+1460}_{-1870}$ &  0.02 $^{+ 0.07}_{- 0.02}$ &  0.71 $^{+ 0.39}_{- 0.71}$ &      946 $^{+    4750}_{-     946}$ &  1.0 \\
HS~0924+3821    & 0.0869 $^{+0.0124}_{-0.0058}$ &      65 $^{+    584}_{-     65}$ &  0.51 $^{+ 0.40}_{- 0.25}$ &  0.04 $^{+ 0.87}_{- 0.04}$ & 12,090 $^{+1700}_{-2220}$ &  0.16 $^{+ 0.02}_{- 0.09}$ &  2.28 $^{+ 1.08}_{- 0.51}$ &        0 $^{+    6330}_{-       0}$ &  0.4 \\
SBS~1152+579    & 0.0924 $^{+0.0117}_{-0.0088}$ &       1 $^{+    119}_{-      1}$ &  0.56 $^{+ 0.39}_{- 0.33}$ &  2.52 $^{+ 0.64}_{- 0.70}$ & 15,190 $^{+2150}_{-2010}$ &  0.21 $^{+ 0.06}_{- 0.06}$ &  0.68 $^{+ 1.09}_{- 0.68}$ &      131 $^{+     512}_{-     131}$ &  3.6 \\
Mrk~209         & 0.0870 $^{+0.0036}_{-0.0052}$ &       1 $^{+    178}_{-      1}$ &  0.68 $^{+ 0.29}_{- 0.35}$ &  0.31 $^{+ 0.50}_{- 0.31}$ & 16,640 $^{+1410}_{-2380}$ &  0.01 $^{+ 0.02}_{- 0.01}$ &  2.43 $^{+ 0.71}_{- 0.94}$ &        0 $^{+      38}_{-       0}$ &  0.5 \\
SBS~0940+544~2  & 0.0923 $^{+0.0068}_{-0.0112}$ &       1 $^{+    120}_{-      1}$ &  0.85 $^{+ 0.48}_{- 0.67}$ &  0.20 $^{+ 0.56}_{- 0.20}$ & 18,320 $^{+1450}_{-2150}$ &  0.03 $^{+ 0.06}_{- 0.03}$ &  2.52 $^{+ 1.30}_{- 1.77}$ &       22 $^{+      94}_{-      22}$ &  0.5 \\
\hline
\multicolumn{9}{c}{Absolute Difference with He~I $\lambda$4026 Removed} \\
\hline
SBS~0917+527    & 0.0054  &       0  &  0.18  &  0.00  &    321  & -0.03  &  0.26  &      235  & - \\
HS~0924+3821    & 0.0018  &     -27  &  0.16  & -0.17  &    494  &  0.00  & -0.01  &        0  & - \\
SBS~1152+579    & 0.0034  &       0  &  0.14  &  0.04  &     16  & -0.02  &  0.24  &       54  & - \\
Mrk~209         & 0.0022  &       0  &  0.31  & -0.18  &    616  &  0.00  &  0.27  &        0  & - \\
SBS~0940+544~2  & 0.0053  &       0  &  0.34  &  0.12  &    -56  & -0.04  &  0.71  &       14  & - \\
Average         & 0.0036                          &      -5                          &  0.23                      & -0.04                      &    278                    & -0.02                      &  0.19                      &       61                            & - \\
\hline
\multicolumn{9}{c}{Fractional Difference with He~I $\lambda$4026 Removed} \\
\hline
SBS~0917+527    & 0.06 &    0.00   &  1.13 &  0.00  &  0.03  & -0.60  &  0.58  &   0.33  & - \\
HS~0924+3821    & 0.02 &   -0.29   &  0.46 & -0.81  &  0.04  &  0.00  & -0.20  &   0.00  & - \\
SBS~1152+579    & 0.04 &    0.00   &  0.33 &  0.02  &  0.00  & -0.09  &  0.55  &   0.70  & - \\
Mrk~209         & 0.03 &    0.00   &  0.84 & -0.37  &  0.04  &  0.00  &  0.13  &   0.00  & - \\
SBS~0940+544~2  & 0.06 &    0.00   &  0.67 &  1.50  &  0.00  & -0.57  &  0.39  &   1.75  & - \\
Average         & 0.04 &   -0.06   &  0.69 &  0.07  &  0.02  & -0.25  &  0.29  &   0.56  & - \\
\hline
\end{tabular}
\caption{Comparison of the solutions for five sample galaxies with and without the inclusion of the He~I $\lambda$4026 observation. Units used in the table are: n$_e$ (cm$^{-3}$), a$_{He}$ and a$_H$ (\AA), T$_e$ (K). Other quantities are dimensionless. }
\label{table:4026}
%\end{sidewaystable}
\end{table}
\end{landscape}

The analyses above indicate the importance of including the faint He~I $\lambda$4026 emission line 
in analyses like ours where the 
underlying absorption is solved for in a minimization.
Solutions lacking it are prone to drifting to larger values of a$_{He}$ and, consequently, y$^{+}$.
As a result of the testing above, and the corresponding behavior manifested in the population of objects without He~I $\lambda$4026, 
these 23 objects without He~I $\lambda$4026 detections 
are excluded from the subsequent analysis and results.  
A primary goal of this work is to reduce systematic bias as much as possible, and the strongest constraint upon a$_{He}$ is provided 
by He~I $\lambda$4026.  As is evident in table \ref{table:4026}, the benefit of an accurate He~I $\lambda$4026 measurement to determining 
the helium abundance should not be underestimated, and thus, requiring it improves the reliability of the sample.
Note that the lack of $\lambda$4026 in other analyses does not automatically result in artificially higher values of y$^{+}$; however,
not properly accounting for underlying helium absorption will certainly affect the reliability of those results.
 
%\section{Cuts on the dataset} \label{Cuts}
\section{Using chi-squared as an analysis tool} \label{Chi}
%\subsection{Using Chi-Squared as an analysis tool} \label{Chi}

\begin{figure}[ht!]
\centering  
\resizebox{\textwidth}{!}{\includegraphics{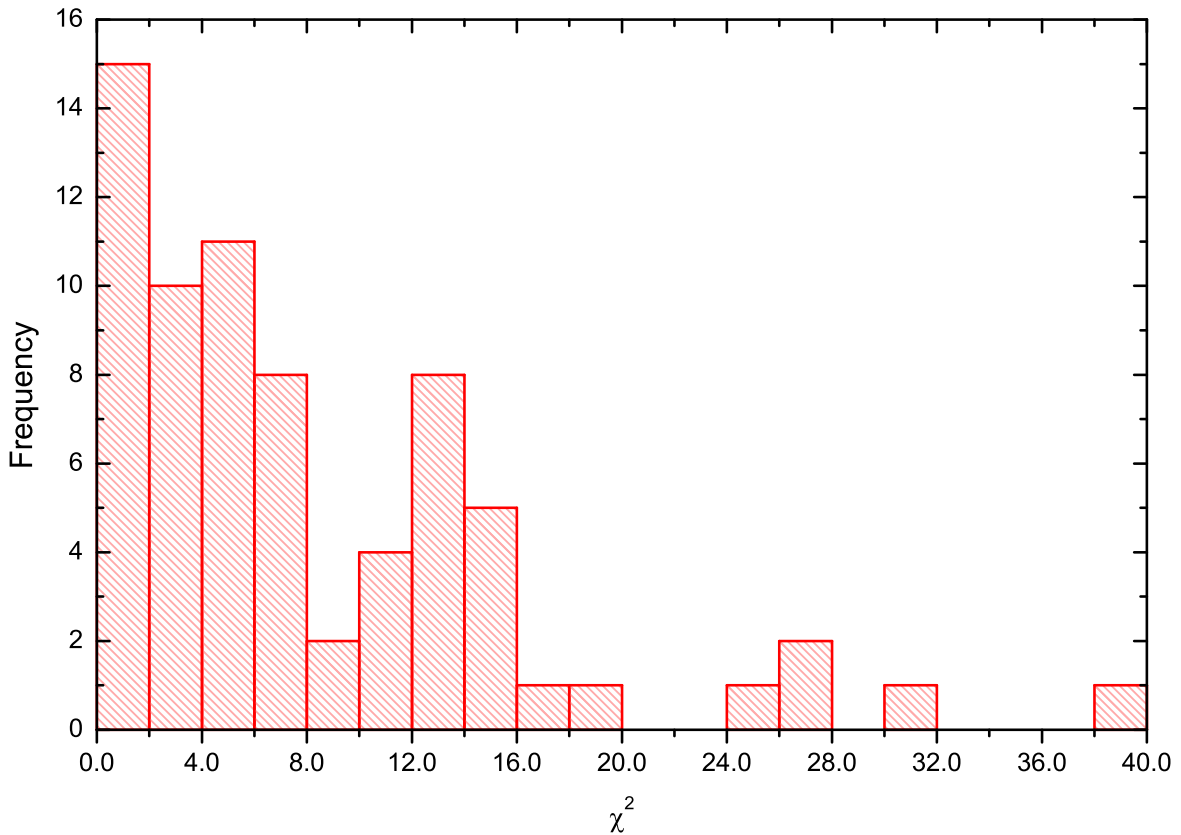}}
\caption{
Histogram of best-fit $\chi^{2}$ for the 70 observations with He~I $\lambda$4026 reported.  25 observations have $\chi^{2}<4$ (95.45\% confidence level).
}
\label{X2}
\end{figure}

After dropping the 23 objects without $\lambda$4026 measurements, 70 objects remain.  Figure \ref{X2} shows the distribution of $\chi^{2}$ values for the solutions for these remaining 70 objects.  Clearly, not all solutions provide good fits to the data (as evidenced by relatively high values of $\chi^{2}$).  Fits with large $\chi^{2}$ may indicate a measurement discrepancy in the line fluxes, an underestimation of the uncertainties, or the possibility that the model used to derive abundances is inappropriate for the object under study.  In either case, points with large $\chi^{2}$ are not reliable, and thus, we exclude them from further analysis in determining Y$_p$.  There are nine observed line ratios used to calculate $\chi^{2}$ and eight model parameters fit to the data; thus, there is only one degree of freedom, modulo correlations\footnote{Correlations between the parameters will increase the effective degrees of freedom.}.  Here we choose to cut the sample at a standard value of $\chi^{2}<4$, corresponding to a 95.45\% confidence level\footnote{We note that of the 23 objects lacking $\lambda$4026, nine had $\chi^2 > 4$ (though, as noted, in the absence of degeneracies among the parameters, we should expect a $\chi^2$ of 0 by fitting 8 parameters to 8 observables.)}.  This cut removes 45 observations, leaving 25 (figure \ref{X2}).  The disappointing result that nearly two-thirds of the sample have solutions with such low likelihoods ($\chi^{2}>4$) is troubling and warrants further investigation.  At this point, we cannot be certain whether this result is due to deficiencies in the model, the observations and their errors, or both.  

Since the number of objects with high values of $\chi^{2}$ is so large, in the following sections, we will carry out parallel analyses in which
we study the reduced sample of 25 objects and compare to the sample of 70.  What we find is that a majority of the solutions with high $\chi^{2}$
values also tend to have questionable values for different physical parameters.  In the end, we conclude that cutting the sample on 
$\chi^{2}$ is fully justified on both theoretical and empirical grounds.   

\section{Investigating problematic points} \label{IllBehaved}

Next, we further investigate the quality of the solutions for the 25 objects meeting the 95\% confidence level.  Clearly it makes sense to exclude non-physical or ambiguous solutions 
even if those solutions have acceptable values of $\chi^{2}$. 
As was demonstrated in AOS2, one of the benefits of an MCMC analysis is that likelihood plots illuminate deficiencies in an object's solution, 
such as very weakly constrained parameters and multiple minima.  
For example, the neutral to ionized hydrogen fraction, $\xi$ is of particular concern. 
As was discovered in AOS, and further investigated in AOS2, $\xi$ is very weakly constrained at relatively low temperatures (higher metallicity), 
and can admit completely unphysical solutions for the relative amount of neutral hydrogen.  

\begin{figure}[ht!]
\centering  
\resizebox{\textwidth}{!}{\includegraphics{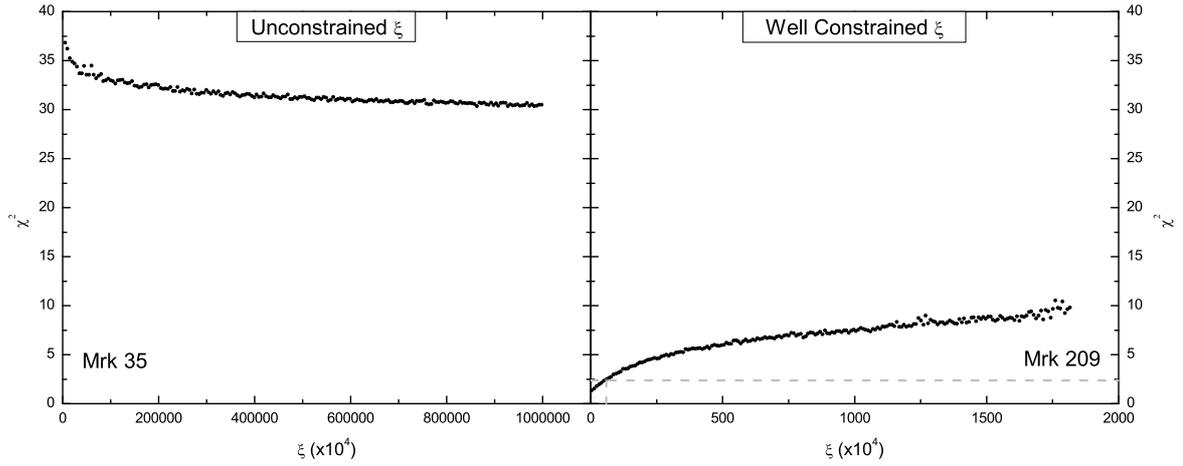}}
\caption{
$\chi^{2}$ versus the neutral to ionized hydrogen fraction, $\xi$, for Mrk~35 and Mrk~209.  On the left, Mrk~35 exhibits an unphysical $\xi$ in contrast to the well determined value for Mrk~209, shown on the right, with the 68\% confidence level marked.   Mrk~35's $\xi$ likelihood plot corresponds to more than 99\% of the hydrogen present being in the neutral state, an unphysical model for an H~II region.
}
\label{Mrk35-Mrk209-HI}
\end{figure}

Figure \ref{Mrk35-Mrk209-HI} illustrates the difference between the well-constrained, physically realistic solution of Mrk~209, and the unbounded solution of Mrk~35.  
For Mrk~35, the global minimum has not yet been reached at a value of $\xi=100$, which corresponds to 99\% of the hydrogen being neutral 
($\xi=10^{-4}$ is a characteristic value for H~II regions).  The best fit solutions with high neutral fraction are not consistent with classifying this 
object as an H~II region (or photo-ionized)!  Observations that do not yield a physically meaningful solution imply either model deficiencies or errors in the observed spectrum.  
Thus, we implement a very conservative criterion that any object whose solution exceeds $\xi=0.333$ (25\% neutral hydrogen) is excluded from further analysis.  
There are two objects from the reduced sample of 25 with large neutral fractions that we exclude with this cut, and thus, we go forward with a reduced sample of 23.  

\begin{figure}[ht!]
\centering  
\resizebox{\textwidth}{!}{\includegraphics{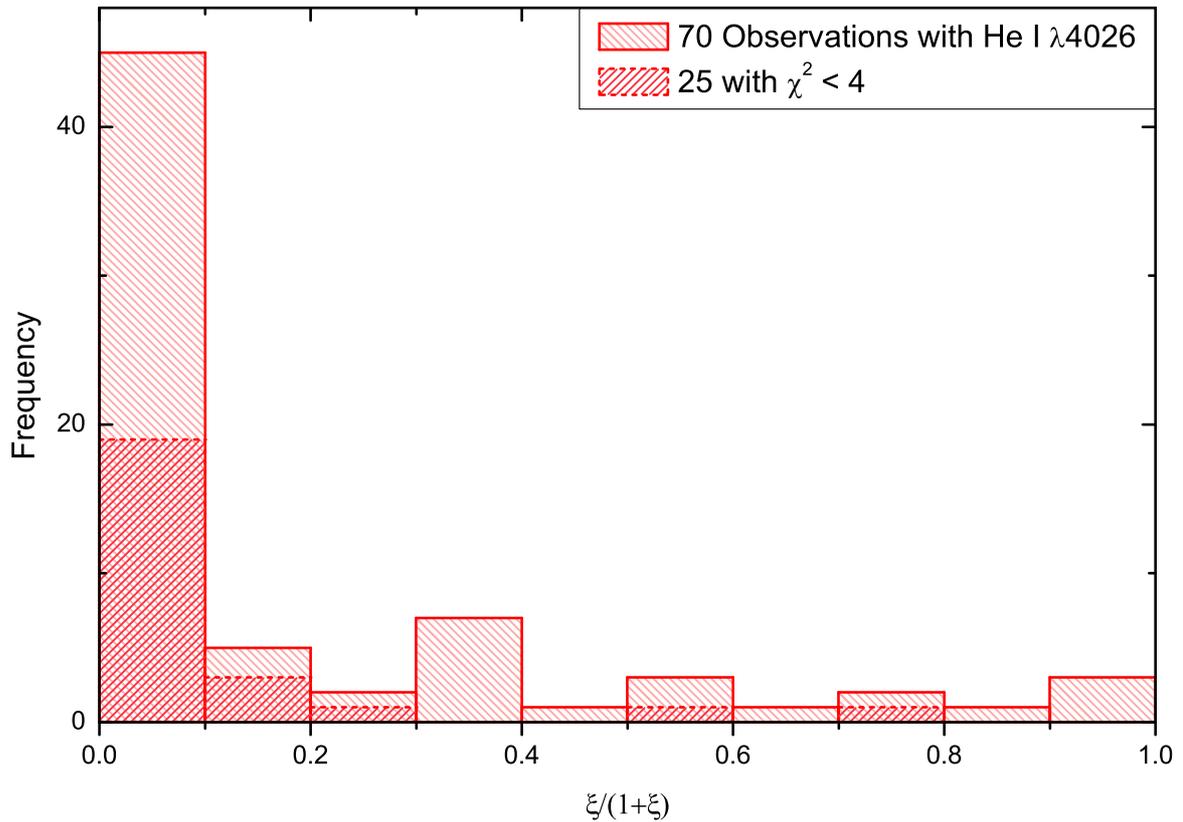}}
\caption{
Histogram of the neutral hydrogen to total hydrogen fraction, $\xi/(1+\xi)$, for the 70 observations with He~I $\lambda$4026.  The 25 with $\chi^{2}<4$ are over-plotted using darker, cross-hatched bars.  %Note that to compress the range of values (0 to 1) for this plot, the solved neutral to ionized hydrogen fraction, $\xi$, is used to calculate the relative contribution of neutral hydrogen.
}
\label{Xi}
\end{figure}

Figure \ref{Xi} shows the distribution of values of $\xi$ in the original sample of 70, prior to the $\chi^2$ cut.  
Note that there are 19 solutions with $\xi>0.333$ in this original sample, 17 of which have 
$\chi^2 > 4$.  Overall, unrealistically large values of the neutral fraction
are indicative of low probability solutions.  
We also identify objects with $\xi>0.01$, and whose lower error bound does not encompass $\xi=0.001$.  
These solutions are flagged for special special consideration and will be discussed later.  There are three such objects in the reduced sample of 23 (seven before the $\chi^2$ cut).

\begin{figure}[ht!]
\centering  
\resizebox{\textwidth}{!}{\includegraphics{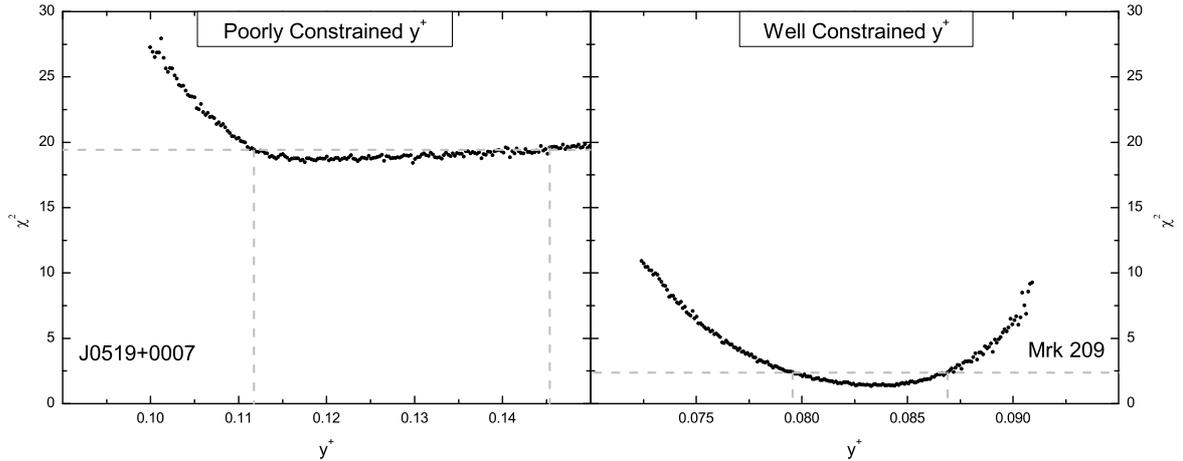}}
\caption{
$\chi^{2}$ versus helium abundance (y$^{+}$) for J0519+0007 and Mrk~209.  On the left, J0519+0007 exhibits a very weakly bounded y$^{+}$ in contrast to the well determined value for Mrk~209, shown on the right, with the 68\% confidence level marked.   J0519+0007's y$^{+}$ likelihood plot is symptomatic of an object whose determination fails on multiple different physical grounds and is completely unreliable.
}
\label{J0519-Mrk209-y}
\end{figure}

Additionally, the likelihood plots of each object are inspected for ambiguous solutions.  Only two objects yielding uncertain solutions are found, both of which have very large $\chi^2$.  SBS~1420+544 exhibits a pronounced double minimum, the lower of which corresponds to a very low temperature ($\sim$10,000 K).  Such behavior calls into question the reliability of this object, and is particularly striking as it is the only object in the dataset to exhibit this phenomenon.  Examples of solutions with double minima were discussed in AOS2.  J0519+0007 is the only object to exhibit a likelihood for y$^{+}$ which is not strongly parabolic about its minimum.  Shown in figure \ref{J0519-Mrk209-y}, the abundance is nearly unbounded for values greater than the minimum, and in fact, the values of y$^{+}$ attained are larger than for any of the other objects.  Mrk~209 is again provided for comparison to a normal, well-behaved abundance.  The values of $\chi^2$ for SBS~1420+544 and J0519+0007 -- 15.7 (for the more physical minimum) and 18.3, respectively -- provide a strong sign of their unreliability.  Similarly, it is interesting that both objects have large optical depth values (4.3 \& 5.6).  Furthermore, J0519+0007 has a temperature much lower than its O[III] temperature (12,800 K vs. 20,700 K) and an unphysically large neutral hydrogen fraction ($\xi$=0.67).  

After removing the objects with problematic solutions we have two samples.  There are 50 objects with well defined solutions, and 23 of these have satisfactorily low values of $\chi^2$.
 
\section{Looking for model outliers} \label{Outliers}

With a dataset comprised of 23 observations, for which the model is a good fit and the parameters are clearly determined, Y$_p$ can be extracted.  However, the models used for correcting for observed flux for optical depth and underlying absorption carry significant systematic uncertainties.  The effect of these systematic uncertainties can be minimized by limiting the size of the corrections.  Additionally, the
solution temperature and the O[III] temperature should be in relatively good agreement for the
solution to make sense.  Here we search for anomalous values of all three physical parameters.
We also discuss the choice of a metallicity baseline.

\subsection{Anomalously large values of optical depth}

\begin{figure}[ht!]
\centering  
\resizebox{\textwidth}{!}{\includegraphics{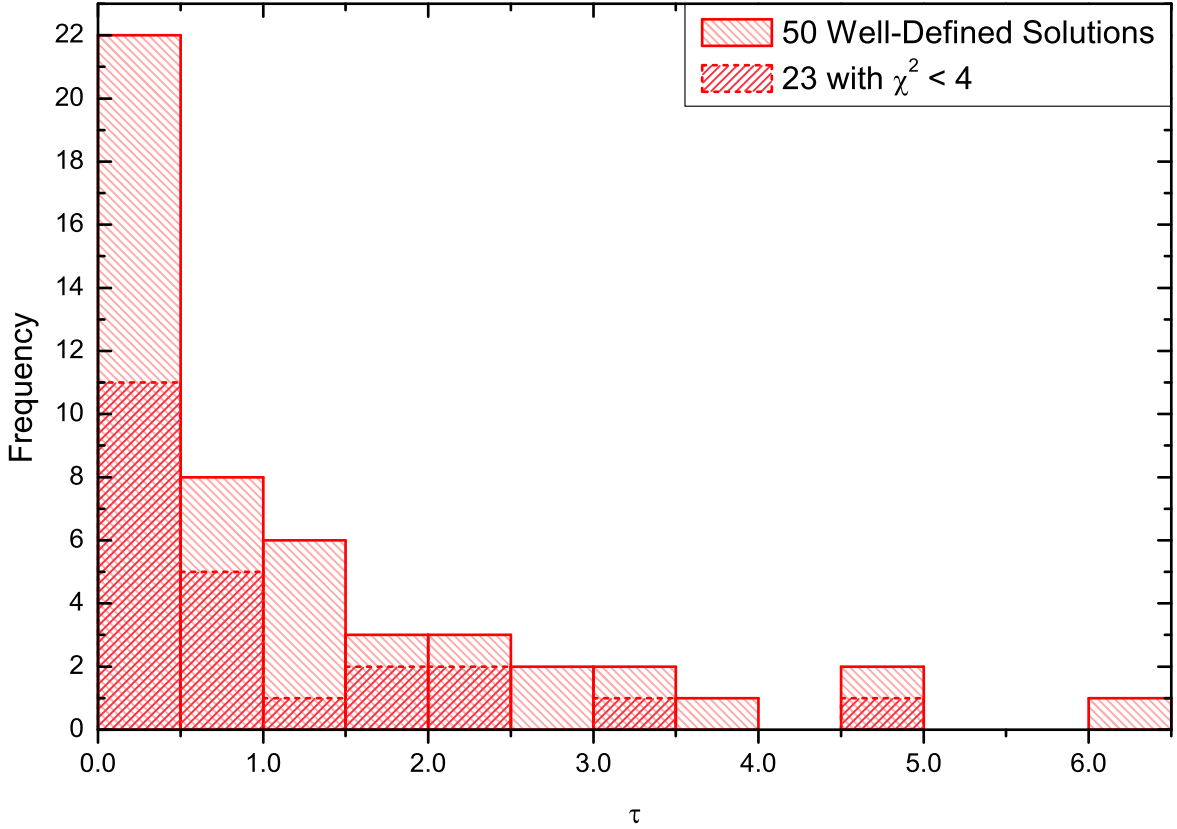}}
\caption{
Histogram of the optical depth, $\tau$, for the 50 well-defined solutions.  The 23 with $\chi^{2}<4$ are over-plotted using darker, cross-hatched bars.
}
\label{Tau}
\end{figure}

In AOS2, it was noted that high values of optical depth increase susceptibility to model deficiencies.  Namely, the radiative transfer calculations of \citet{bss02} do not take into account any expansion, non-uniformity, or turbulence in the H~II region.  A histogram showing the distribution of the optical depth in objects before and after the $\chi^2<4$ cut is shown in figure \ref{Tau}.  There are originally 3 potential outliers with $\tau>4$.  Reassuringly though, the $\chi^2$ cut removes two of these.  The remaining object is flagged to ascertain the impact of questionable points on the regression.

\subsection{Anomalously large values of underlying absorption}

\begin{figure}[ht!]
\centering  
\resizebox{\textwidth}{!}{\includegraphics{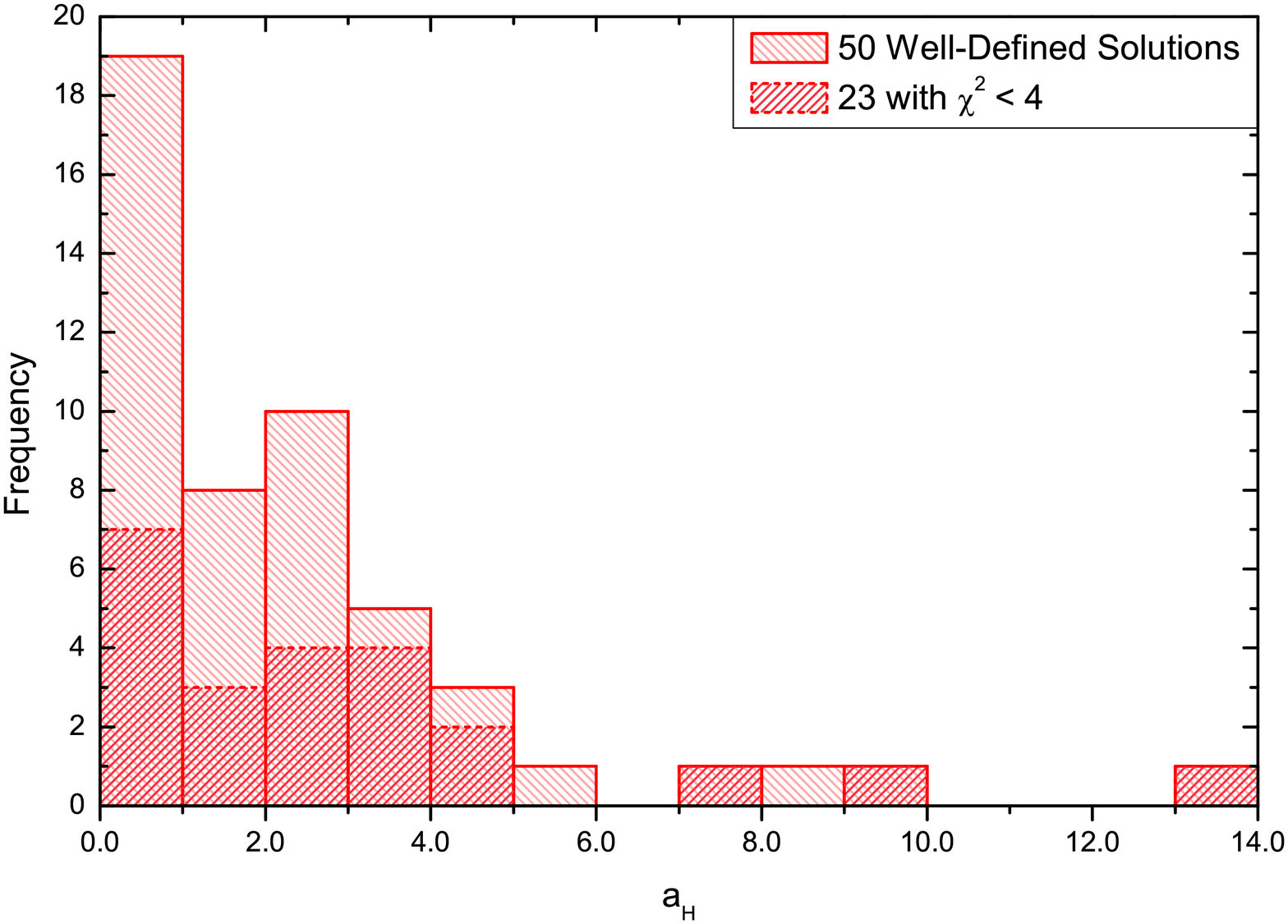}}
\caption{
Histogram of the underlying hydrogen absorption, a$_H$, for the 50 well-defined solutions.  The 23 with $\chi^{2}<4$ are over-plotted using darker, cross-hatched bars.
}
\label{AH}
\end{figure}

\begin{figure}[ht!]
\centering  
\resizebox{\textwidth}{!}{\includegraphics{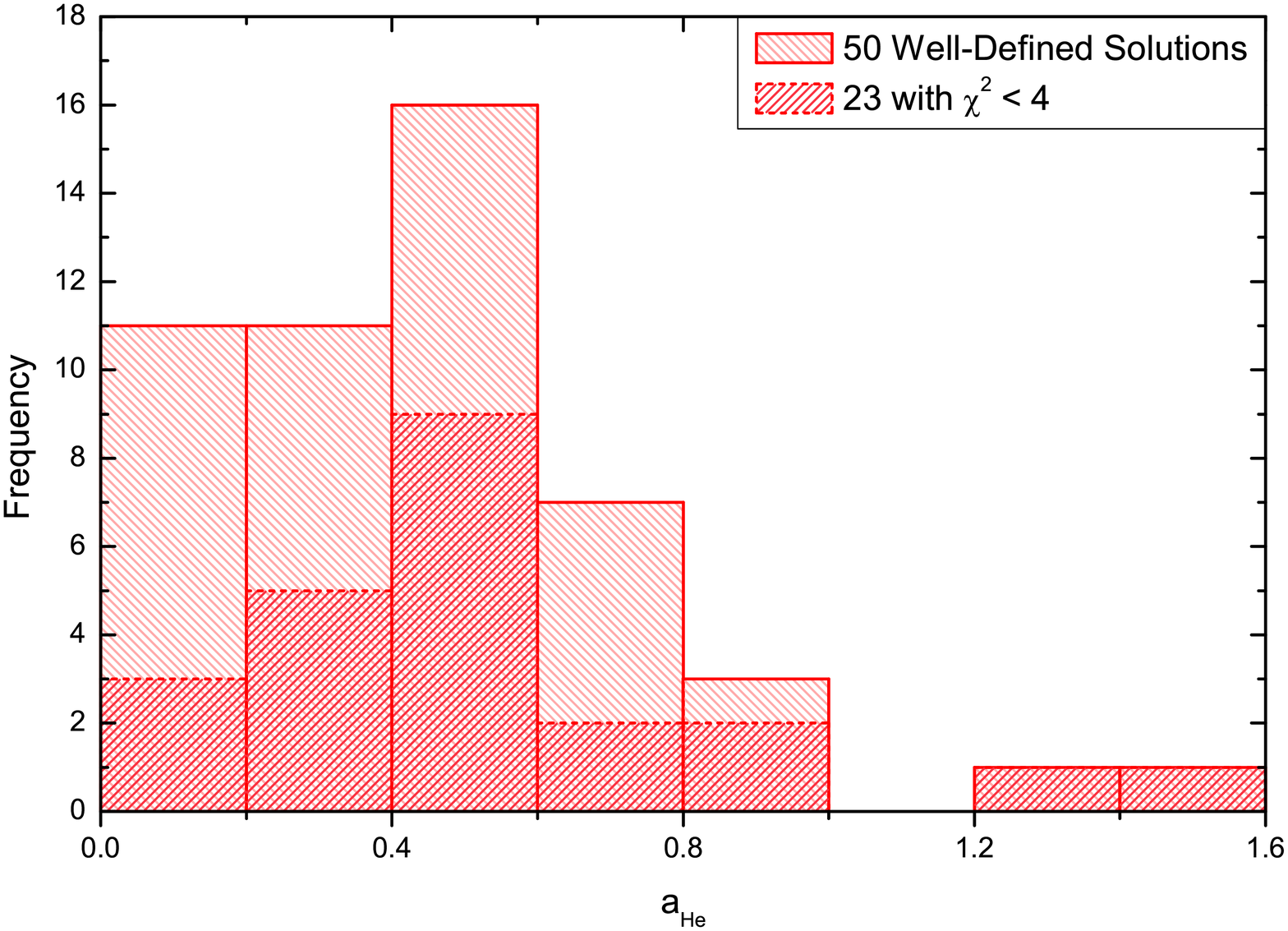}}
\caption{
Histogram of the underlying hydrogen absorption, a$_He$, for the 50 well-defined solutions.  The 23 with $\chi^{2}<4$ are over-plotted using darker, cross-hatched bars.
}
\label{AHe}
\end{figure}

Next, a similar investigation is performed for underlying absorption.  Though AOS added wavelength dependence to the modeling of underlying absorption, it is still a difficult effect to accurately quantify.  It was for this reason that \citet{os04} only analyzed galaxies with $W(H\beta) > 200$ \AA, which minimizes the effect of underlying absorption.  The histograms for a$_{H}$ and a$_{He}$ are
shown in figures \ref{AH} and \ref{AHe}, respectively. As one can see, there are four potential outliers with a$_H > 6$ \AA\  and two with a$_{He} > 1$ \AA.   One of the objects is anomalously high in both, and one is removed by the $\chi^2$ selection.  Thus, four objects are flagged for further analysis with acceptable values of $\chi^2$ but anomalous solution parameters.

\subsection{Anomalous values of temperature}

The comparison of the solution temperature and the O[III] temperature provides a final physical parameter to investigate here.  In this case, the issue is not one of outliers and their model realizability, but the physical constraint on these two temperatures:  they should be close to one another, with T(O~III) serving as a loose upper bound for T.  Note that T is very weakly constrained by T(O~III), but primarily determined by the helium lines (i.e., T$\sim$T(He~II); see AOS2 for further discussion).

\begin{figure}[ht!]
\centering  
\resizebox{\textwidth}{!}{\includegraphics{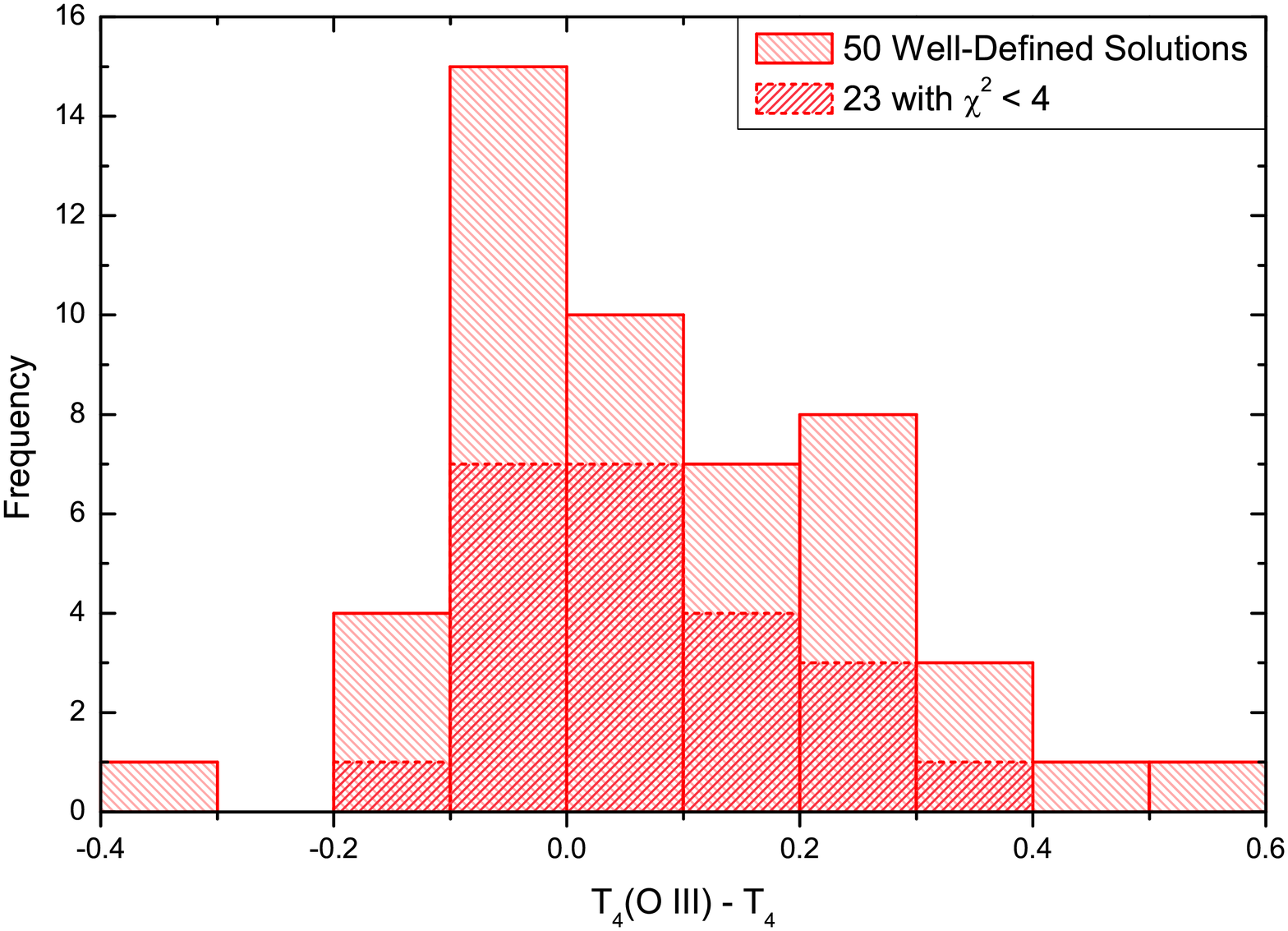}}
\caption{
Histogram of the difference in the O[III] temperature and solution temperature, $T_{4}(O~III) - T_{4}$, 
where $T_{4}$ is in units of 10$^4$K, for the 50 well-defined solutions.  The 23 with $\chi^{2}<4$ are over-plotted using darker, cross-hatched bars.  It is expected that T(O~III) should be greater than T, and that T should closely track T(O~III).  That some values of T will be larger than T(O~III) is a result of the combination of the close relationship of their actual physical values and the relatively large uncertainty on T.  However, the most extreme values are not easily justified.  As result, it is not surprising that the $\chi^{2}$ cut removes most of them.
}
\label{TOIII}
\end{figure}

Figure \ref{TOIII} shows the difference, $\Delta T = T(O~III) - T$, before and after the $\chi^2$ cut.  Since, in the presence of temperature fluctuations, the O[III] temperature is biased toward higher values,
T(He~II) is estimated to be less than T(O~III) by 3-11\% (see AOS2).  Additionally, the uncertainties on T are very large, averaging $\sim$2000 K over the set of 50 observations.  As a result, there will be a spread in $\Delta T$, including negative values.  However, the most extreme values for this difference, $\Delta T < -3000$ K and $\Delta T > 5000$ K are very unlikely results (as is the number of results with $\Delta T < 0$ K).  Reinforcing the utility of the $\chi^2$ selection, the $\chi^2<4$ subset does not admit the extreme values, and the distribution now reflects the physical expectation.

\subsection{Choosing a baseline in metallicity}

\begin{figure}[ht!]
\centering  
\resizebox{\textwidth}{!}{\includegraphics{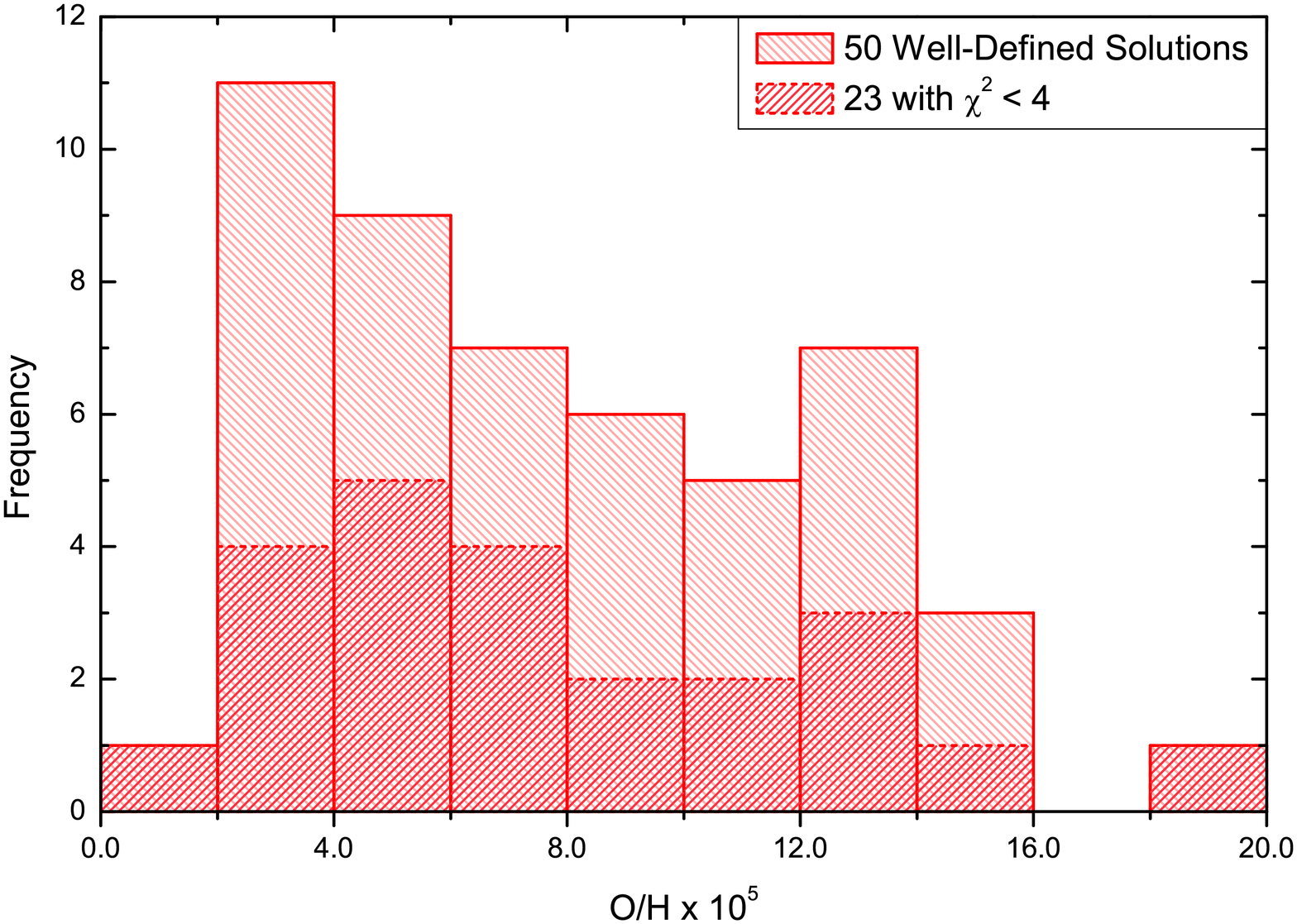}}
\caption{
Histogram of the oxygen to hydrogen ratio, O/H, for the 50 well-defined solutions.  The 23 with $\chi^{2}<4$ are over-plotted using darker, cross-hatched bars.
}
\label{OH2}
\end{figure}

The last criterion for dataset selection is the metallicity baseline.  Extending the metallicity baseline to higher values increases the systematic uncertainty in the value of the 
primordial helium abundance which is due to the assumption of a linear relationship of He/H with chemical evolution.  
For this reason, a shorter metallicity baseline is preferred, but there is a trade-off between a metallicity cut and sample size.
Clearly it is not desirable to truncate the dataset significantly and thus lose precision on the determination of Y$_p$.  
Figure \ref{OH2} shows that both samples exhibit a reasonable coverage of points up through $O/H = 15.1 \times 10^{-5}$ and then a possible outlier at $18.1 \times 10^{-5}$.  
This last point could have undue leverage on the regression if included and is therefore excluded.  
The exclusion of the largest metallicity point yields a final dataset of 22 objects with well-defined solutions that are good fits to the data.  

A summary of the cuts and flagged objects on the sample is presented in table \ref{table:Cuts}.  To emphasize the utility of the $\chi^2$ goodness-of-fit test, the sequence is provided with $\chi^2<4$ applied first, as was presented in the above description, and after the other criteria.  The resulting final dataset is the same, but the effectiveness of $\chi^2$ for identifying problematic solutions is apparent.

\begin{table}[ht!]
\centering
\vskip .1in
\begin{tabular}{lrr}
%\tabletypesize{\footnotesize}
%\tablewidth{0pt}
\hline\hline
Criterion				& $\chi^{2}<4$ first	 & if $\chi^{2}<4$ reserved\\
					&			 & until last\\
\hline
HeBCD (ITS07)				& 93	 & 93 \\
\hline
\underline{\textit{Cuts}} \\
He~I $\lambda$4026 Not Detected		& 23/93	 & 23/93 \\
$\chi^{2}>4$		   		& 45/70  &  - \\
$\xi>0.333$				&  2/25	 & 19/70 \\
Degenerate Solution			&  0/23	 &  1/51 \\
\hline
Subtotal:  Well-Defined Solutions	& 23	 & 50 \\
\hline
\underline{\textit{Flagged}} \\
%\hline
\hspace{0.1in} $\xi>0.01$			&  3     &  7 \\
\hspace{0.1in} $\tau>4$				&  1   	 &  3 \\
\hspace{0.1in} a$_H > 6$ \AA			&  3     &  4 \\
\hspace{0.1in} a$_{He} > 1$ \AA			&  2     &  2 \\
\hspace{0.1in} $T(O~III) - T < -3000~K$		&  0     &  1 \\
\hspace{0.1in} $T(O~III) - T >  5000~K$		&  0     &  1 \\
\hline
Subtotal:  Unique Flagged		&  8	 & 17 \\
\hline
\underline{\textit{Cuts}} \\
$O/H \ge 15.2 \times 10^{-5}$		&  1/23	 &  1/50 \\
$\chi^{2}>4$		   		&  -     & 27/49 \\
\hline
Final Dataset				& 22	 & 22 \\
\hspace{0.1in} Flagged			&  8	 &  8 \\
\hspace{0.1in} Qualifying		& 14	 & 14 \\
\hline
\end{tabular}
\caption{Breakdown of the cuts and flags on the sample.  To better highlight the behavior of the physical parameters and emphasize the role of $\chi^2$, each criterion is tabulated for the case where the cut on $\chi^2$ is made first and the case where it is made afterward.}
\label{table:Cuts}
\end{table}

\section{Results from the Final Dataset} \label{Results}

\begin{figure}[ht!]
\centering  
\resizebox{\textwidth}{!}{\includegraphics{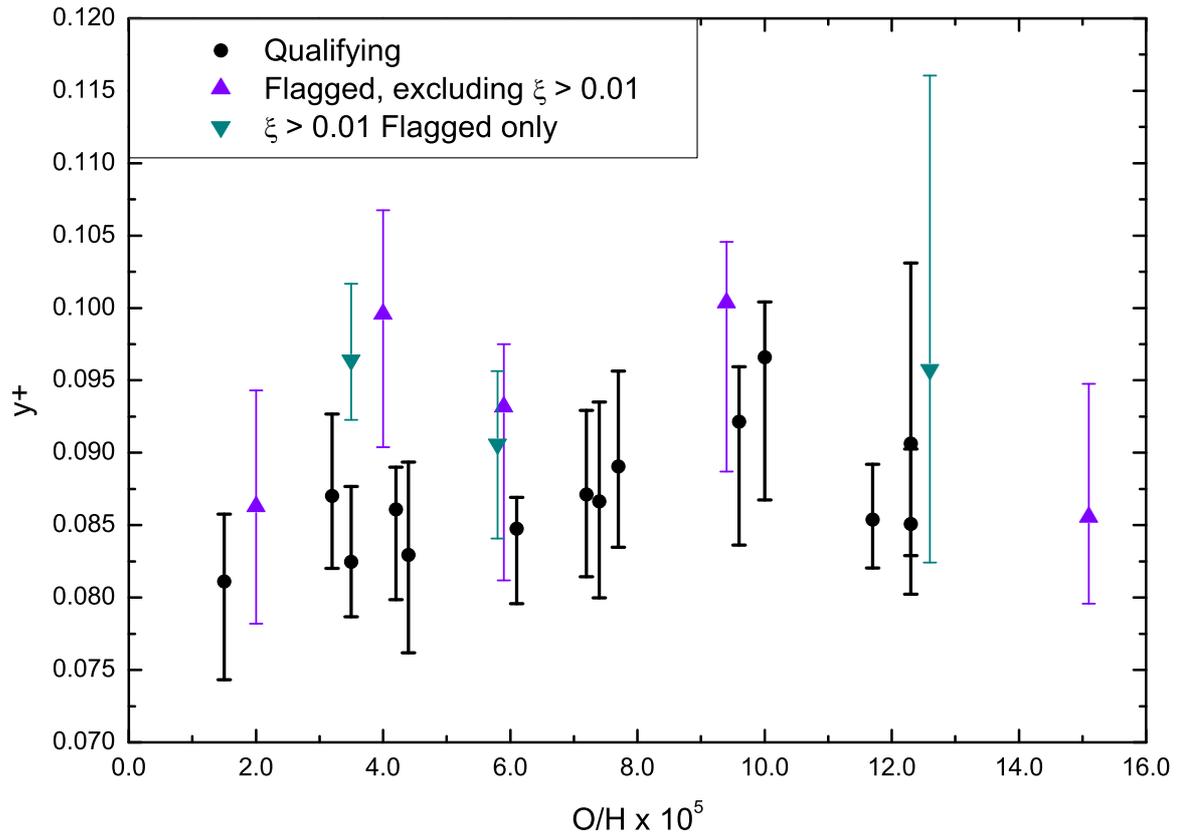}}
\caption{
Plot of y$^{+}$ vs O/H for the 22 objects meeting the prescribed reliability standards.  The upward triangles signify points flagged for large outlier values in optical depth or underlying absorption.  The downward triangles signify points flagged for large neutral hydrogen fractions. 
}
\label{y+OH}
\end{figure}

The 22 objects for which the model is a good fit, which return physically meaningful parameter solutions, and which provide a robust metallicity baseline for regression comprise the Final Dataset, which we use to determine Y$_p$.  
Each of the objects, with the results of their best-fit solutions and uncertainties, are presented in table \ref{table:GTO}.  
Figure \ref{y+OH} presents the derived y$^{+}$ values as a function of O/H.  
The seven objects flagged for large outlier values of $\tau$, a$_H$, a$_He$, and $\xi$ are highlighted with different symbols in figure \ref{y+OH}.  These flagged data points show a possible systematic shift to larger values of y$^{+}$; the average value of y$^{+}$ for the flagged objects is 9\% higher than that of the unflagged objects in the Final Dataset.

\begin{landscape}
\begin{table}[b!]\footnotesize
%\begin{sidewaystable}[b]\footnotesize
\centering
\vskip .1in
\begin{tabular}{lccccccccc}
%\tabletypesize{\footnotesize}
%\tablewidth{0pt}
\hline
\hline
Object 			& He$^+$/H$^+$     		   &  n$_e$     		      & a$_{He}$     		   & $\tau$     			& T$_e$     		   & C(H$\beta$)     	  	& a$_H$		     & $\xi$ $\times$ 10$^4$ 			& $\chi^2$ \\
\hline
% &&& Final Dataset Not Flagged \\
\multicolumn{10}{c}{Final Dataset Not Flagged} \\
\hline
I~Zw~18SE~1        	& 0.0811 $^{+0.0046}_{-0.0068}$ &       4 $^{+    248}_{-      4}$ &  0.29 $^{+ 0.23}_{- 0.24}$ &  0.43 $^{+ 0.70}_{- 0.43}$ & 18,190 $^{+2500}_{-2960}$ &  0.01 $^{+ 0.02}_{- 0.01}$ &  3.80 $^{+ 0.62}_{- 0.69}$ &        0 $^{+      26}_{-       0}$ &  0.5 \\
SBS~0940+544~2    	& 0.0870 $^{+0.0057}_{-0.0050}$ &       8 $^{+    131}_{-      8}$ &  0.51 $^{+ 0.18}_{- 0.18}$ &  0.08 $^{+ 0.64}_{- 0.08}$ & 18,380 $^{+1330}_{-2500}$ &  0.07 $^{+ 0.03}_{- 0.05}$ &  1.81 $^{+ 1.34}_{- 1.09}$ &        8 $^{+      74}_{-       8}$ &  0.7 \\
Tol~65           	& 0.0825 $^{+0.0052}_{-0.0038}$ &     211 $^{+    305}_{-    135}$ &  0.67 $^{+ 0.11}_{- 0.08}$ &  0.00 $^{+ 0.47}_{- 0.00}$ & 17,330 $^{+1620}_{-2910}$ &  0.13 $^{+ 0.02}_{- 0.05}$ &  3.54 $^{+ 0.82}_{- 0.60}$ &        1 $^{+     137}_{-       1}$ &  2.4 \\
SBS~1415+437~(No.~2)	& 0.0861 $^{+0.0029}_{-0.0063}$ &       5 $^{+    343}_{-      5}$ &  0.54 $^{+ 0.07}_{- 0.12}$ &  1.16 $^{+ 0.75}_{- 1.16}$ & 15,440 $^{+1910}_{-2320}$ &  0.00 $^{+ 0.02}_{- 0.00}$ &  3.48 $^{+ 0.72}_{- 0.72}$ &        0 $^{+      57}_{-       0}$ &  1.9 \\
HS~1442+4250     	& 0.0830 $^{+0.0064}_{-0.0068}$ &       1 $^{+    257}_{-      1}$ &  0.46 $^{+ 0.40}_{- 0.38}$ &  0.29 $^{+ 0.85}_{- 0.29}$ & 14,880 $^{+2180}_{-2650}$ &  0.03 $^{+ 0.03}_{- 0.03}$ &  0.00 $^{+ 1.15}_{- 0.00}$ &      127 $^{+    1550}_{-     127}$ &  2.2 \\
Mrk~209          	& 0.0848 $^{+0.0022}_{-0.0052}$ &       1 $^{+    293}_{-      1}$ &  0.37 $^{+ 0.11}_{- 0.16}$ &  0.49 $^{+ 0.41}_{- 0.49}$ & 16,030 $^{+1380}_{-2580}$ &  0.01 $^{+ 0.02}_{- 0.01}$ &  2.16 $^{+ 0.95}_{- 0.80}$ &        0 $^{+      59}_{-       0}$ &  1.4 \\
Mrk~71~(No.~1)           & 0.0871 $^{+0.0058}_{-0.0057}$ &      93 $^{+    341}_{-     93}$ &  0.60 $^{+ 0.28}_{- 0.25}$ &  1.75 $^{+ 0.67}_{- 0.83}$ & 14,170 $^{+1860}_{-2090}$ &  0.09 $^{+ 0.04}_{- 0.05}$ &  1.13 $^{+ 1.71}_{- 1.13}$ &      132 $^{+    1270}_{-     132}$ &  1.2 \\
SBS~0917+527     	& 0.0866 $^{+0.0069}_{-0.0067}$ &       1 $^{+    196}_{-      1}$ &  0.16 $^{+ 0.13}_{- 0.12}$ &  0.00 $^{+ 0.59}_{- 0.00}$ & 12,770 $^{+1450}_{-1830}$ &  0.05 $^{+ 0.05}_{- 0.05}$ &  0.45 $^{+ 0.65}_{- 0.45}$ &      711 $^{+    5510}_{-     711}$ &  1.2 \\
SBS~1152+579     	& 0.0891 $^{+0.0066}_{-0.0056}$ &       1 $^{+    130}_{-      1}$ &  0.42 $^{+ 0.15}_{- 0.14}$ &  2.48 $^{+ 0.75}_{- 0.61}$ & 15,170 $^{+1840}_{-2040}$ &  0.23 $^{+ 0.04}_{- 0.05}$ &  0.44 $^{+ 0.81}_{- 0.44}$ &       77 $^{+     498}_{-      77}$ &  3.8 \\
SBS~1054+365     	& 0.0921 $^{+0.0038}_{-0.0085}$ &       1 $^{+    428}_{-      1}$ &  0.49 $^{+ 0.18}_{- 0.24}$ &  0.51 $^{+ 0.57}_{- 0.51}$ & 12,040 $^{+1720}_{-1620}$ &  0.00 $^{+ 0.07}_{- 0.00}$ &  2.87 $^{+ 0.49}_{- 1.04}$ &     2040 $^{+    5160}_{-    2040}$ &  1.0 \\
SBS~0926+606     	& 0.0966 $^{+0.0038}_{-0.0099}$ &       1 $^{+    296}_{-      1}$ &  0.70 $^{+ 0.13}_{- 0.20}$ &  0.12 $^{+ 0.61}_{- 0.12}$ & 12,790 $^{+1520}_{-1750}$ &  0.00 $^{+ 0.07}_{- 0.00}$ &  0.75 $^{+ 0.33}_{- 0.75}$ &     1460 $^{+    4470}_{-    1460}$ &  0.7 \\
SBS~1135+581     	& 0.0854 $^{+0.0038}_{-0.0034}$ &     468 $^{+    913}_{-    468}$ &  0.44 $^{+ 0.09}_{- 0.08}$ &  0.00 $^{+ 0.59}_{- 0.00}$ & 11,270 $^{+1880}_{-1660}$ &  0.10 $^{+ 0.02}_{- 0.03}$ &  3.42 $^{+ 0.47}_{- 0.54}$ &        0 $^{+    3160}_{-       0}$ &  2.7 \\
HS~0924+3821     	& 0.0851 $^{+0.0052}_{-0.0049}$ &      92 $^{+    836}_{-     92}$ &  0.35 $^{+ 0.13}_{- 0.12}$ &  0.21 $^{+ 0.82}_{- 0.21}$ & 11,600 $^{+1480}_{-2190}$ &  0.16 $^{+ 0.02}_{- 0.05}$ &  2.29 $^{+ 0.71}_{- 0.52}$ &        0 $^{+   10100}_{-       0}$ &  0.9 \\
UM~439           	& 0.0906 $^{+0.0125}_{-0.0078}$ &     185 $^{+    660}_{-    185}$ &  0.01 $^{+ 0.24}_{- 0.01}$ &  3.29 $^{+ 0.95}_{- 1.01}$ & 14,580 $^{+2040}_{-3290}$ &  0.20 $^{+ 0.05}_{- 0.07}$ &  0.57 $^{+ 1.16}_{- 0.57}$ &      218 $^{+    2980}_{-     218}$ &  2.9 \\
%UM~396           	& 0.0834 $^{+0.0053}_{-0.0036}$ &     702 $^{+   1470}_{-    491}$ &  0.00 $^{+ 0.08}_{- 0.00}$ &  0.15 $^{+ 0.85}_{- 0.15}$ & 12240 $^{+1660}_{-2180}$ &  0.28 $^{+ 0.02}_{- 0.06}$ &  4.78 $^{+ 1.14}_{- 0.78}$ &        0 $^{+    4200}_{-       0}$ &  2.3 \\
\hline
% &&& Final Dataset with Flags \\
\multicolumn{10}{c}{Final Dataset with Flags} \\
\hline
SBS~0335-052E1   	& 0.0863 $^{+0.0080}_{-0.0081}$ &       4 $^{+    182}_{-      4}$ &  0.24 $^{+ 0.18}_{- 0.20}$ &  4.96 $^{+ 0.89}_{- 1.01}$ & 18,120 $^{+2330}_{-2680}$ &  0.09 $^{+ 0.05}_{- 0.06}$ &  1.51 $^{+ 1.21}_{- 1.25}$ &       17 $^{+     126}_{-      17}$ &  0.9 \\
HS~0122+0743     	& 0.0996 $^{+0.0072}_{-0.0092}$ &       1 $^{+     97}_{-      1}$ &  1.40 $^{+ 0.21}_{- 0.23}$ &  0.55 $^{+ 0.50}_{- 0.55}$ & 18,070 $^{+2270}_{-1790}$ &  0.06 $^{+ 0.06}_{- 0.06}$ &  4.34 $^{+ 1.47}_{- 1.57}$ &       38 $^{+     109}_{-      38}$ &  1.1 \\
UM~461           	& 0.0932 $^{+0.0043}_{-0.0120}$ &       1 $^{+    295}_{-      1}$ &  0.90 $^{+ 0.27}_{- 0.35}$ &  2.07 $^{+ 0.67}_{- 0.97}$ & 18,340 $^{+2110}_{-2850}$ &  0.00 $^{+ 0.07}_{- 0.00}$ &  7.72 $^{+ 0.87}_{- 1.94}$ &       28 $^{+      98}_{-      28}$ &  1.5 \\
HS~0811+4913     	& 0.1004 $^{+0.0042}_{-0.0117}$ &      26 $^{+    323}_{-     26}$ &  1.39 $^{+ 0.25}_{- 0.46}$ &  0.99 $^{+ 0.52}_{- 0.99}$ & 14,850 $^{+1750}_{-2070}$ &  0.01 $^{+ 0.07}_{- 0.01}$ &  9.13 $^{+ 0.84}_{- 2.89}$ &      396 $^{+    1370}_{-     396}$ &  1.5 \\
UM~238           	& 0.0856 $^{+0.0092}_{-0.0060}$ &    1150 $^{+   3700}_{-    624}$ &  0.92 $^{+ 0.62}_{- 0.46}$ &  0.53 $^{+ 0.98}_{- 0.53}$ & 12911 $^{+1950}_{-2770}$ &  0.24 $^{+ 0.04}_{- 0.08}$ & 13.21 $^{+ 3.69}_{- 2.19}$ &      178 $^{+    4280}_{-     178}$ &  0.9 \\
UGC~4483         	& 0.0964 $^{+0.0053}_{-0.0041}$ &       1 $^{+    113}_{-      1}$ &  0.47 $^{+ 0.09}_{- 0.08}$ &  0.36 $^{+ 0.56}_{- 0.36}$ & 14,090 $^{+1600}_{-1680}$ &  0.09 $^{+ 0.03}_{- 0.04}$ &  0.00 $^{+ 0.54}_{- 0.00}$ &      716 $^{+    2070}_{-     525}$ &  2.2 \\
SBS~1331+493     	& 0.0906 $^{+0.0051}_{-0.0065}$ &     138 $^{+    507}_{-    138}$ &  0.23 $^{+ 0.10}_{- 0.13}$ &  0.97 $^{+ 0.58}_{- 0.97}$ & 12,910 $^{+1940}_{-1570}$ &  0.00 $^{+ 0.04}_{- 0.00}$ &  0.29 $^{+ 0.76}_{- 0.29}$ &     1440 $^{+    3720}_{-    1170}$ &  3.8 \\
SBS~1533+574B    	& 0.0957 $^{+0.0203}_{-0.0133}$ &     401 $^{+    924}_{-    401}$ &  0.46 $^{+ 0.20}_{- 0.20}$ &  1.53 $^{+ 1.44}_{- 1.32}$ & 12,620 $^{+2750}_{-1460}$ &  0.27 $^{+ 0.11}_{- 0.10}$ &  2.56 $^{+ 0.95}_{- 1.03}$ &     2630 $^{+    3690}_{-    2450}$ &  2.7 \\
\hline
\end{tabular}
\caption{Physical Conditions and He$^+$/H$^+$ Abundance Solutions of Final Dataset}
\label{table:GTO}
%\end{sidewaystable}
\end{table}
\end{landscape}

The primary goal of this work, the primordial helium abundance (mass fraction), Y$_{p}$, can now be calculated for several subsets of the final dataset.  A regression of Y, the helium mass fraction, versus O/H, the oxygen to hydrogen mass fraction, is used to extrapolate to the primordial value\footnote{This work takes $Z=20(O/H)$ such that $Y=\frac{4y(1-20(O/H))}{1+4y}$}.  The O/H values are taken directly from ITS07.  

\begin{figure}
\resizebox{\textwidth}{!}{\includegraphics{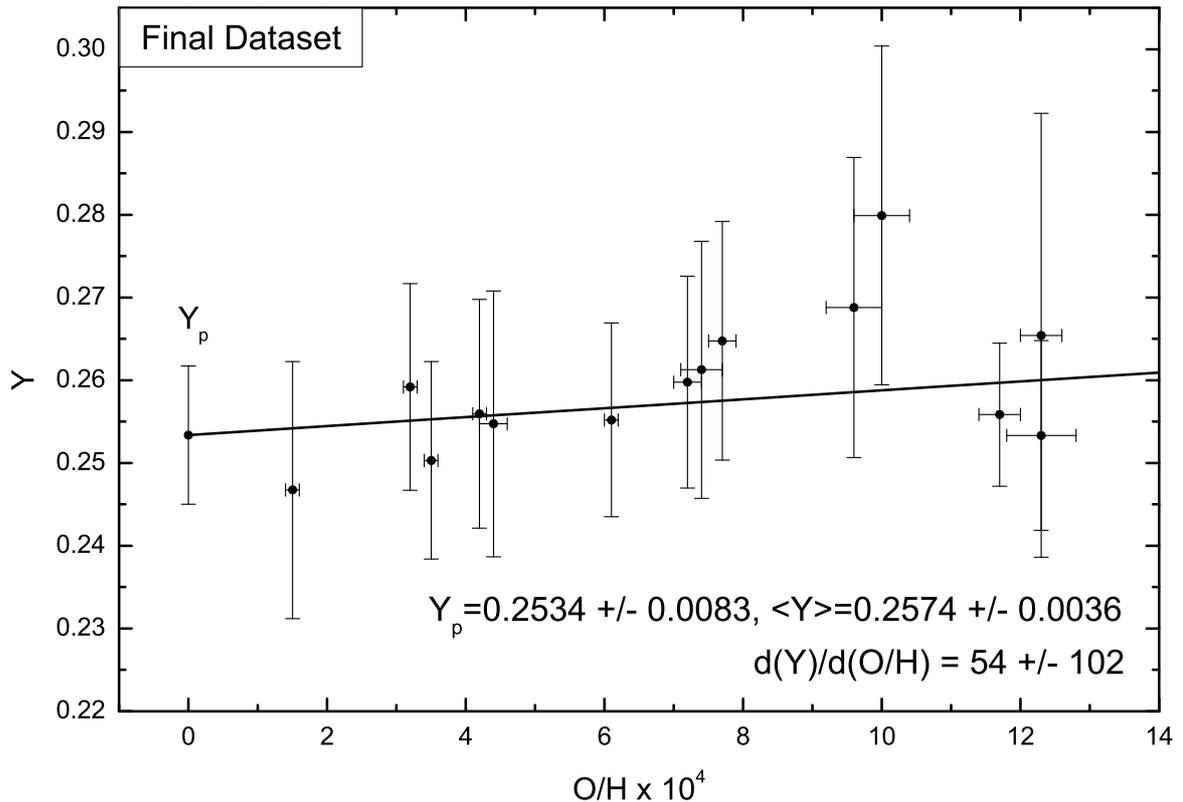}}
\caption{
Helium abundance (mass fraction) versus oxygen to hydrogen ratio regression calculating the primordial helium abundance.
}
\label{Y_OH}
\end{figure}

Because it minimizes confounding systematic effects, our preferred dataset is the 14 qualifying points.  
The relevant values for its regression are given in table \ref{table:PH}.  The regression yields,
\beq
Y_p = 0.2534 \pm 0.0083,
\label{eq:Yp}
\eeq
with a slope of 54 $\pm$ 102 and a $\chi^2$ of 2.9.  The result is shown in figure \ref{Y_OH}.  
Note that the expected value of $\chi^2$ for this dataset is $\sim$ 12, so the resultant $\chi^2$ is unexpectedly low.
This result for $Y_p$ agrees well with the WMAP value of $Y_p = 0.2487 \pm 0.0002$. 
AOS2 determined $Y_p = 0.2609 \pm 0.0117$ ($0.2573^{+0.0033}_{-0.0088}$ with the slope restricted to be positive).  
Given their large uncertainties, these results are in agreement with the newer result.  
The smaller uncertainty on the unconstrained fit is a direct result of the increased sample size.
%improved, slightly lower value of Y$_p$ found in this work may be a result of the more stringent screening for systematic effects.  

As the O/H domain is limited, an estimate of Y$_p$ using the mean value is justified and gives,
\beq
Y_p = 0.2574 \pm 0.0036.
\eeq
This is not significantly different from the result of the regression fit; however, the uncertainty is decreased by more than a factor of two.  
%It will take a substantial observational effort to increase the sample size of acceptable objects in order to decrease the uncertainty in the regression fit to the level of that of the straight average.

Including the flagged objects raises the intercept and reduces error to $0.2611 \pm 0.0067$ with a slope of 0 $\pm$ 86.  
The reduced uncertainty is a result of the increased number of points in the regression, and the possible systematic bias toward larger y$^{+}$ within the flagged dataset raises the intercept.  \citet{os04} restricted the metallicity baseline to $O/H = 9.2 \times 10^{-5}$.  
Adopting the same metallicity cut with the dataset of this work decreases the intercept substantially to $0.2465 \pm 0.0134$ and produces a strongly positive slope (though still consistent with 0) of 196 $\pm$ 230.  Using all 93 observations included in their HeBCD sample, ITS07 determined $Y_p = 0.2516 \pm 0.0011$.  Their much smaller uncertainty is achieved primarily though the use of the full sample of observations.  Table \ref{table:Yp's} summarizes the calculated regression Y$_p$ and slope, as well as the mean, $<Y>$, 
for several subsets of the Final Dataset found in this work.

\begin{table}[ht!]
\centering
\vskip .1in
\begin{tabular}{lcccc}
%\tabletypesize{\footnotesize}
%\tablewidth{0pt}
\hline\hline
Object & 	He$^+$/H$^+$ 	      & He$^{++}$/H$^+$     & Y 		  & O/H $\times$ 10$^5$ \\
\hline
I~Zw~18SE~1 		&	0.08112	$\pm$	0.00680	&	0.0008	$\pm$	0.0008	&	0.2467	$\pm$	0.0155	&	1.5	$\pm$	0.1	\\
SBS~0940+544~2 		&	0.08701	$\pm$	0.00567	&	0.0005	$\pm$	0.0005	&	0.2592	$\pm$	0.0125	&	3.2	$\pm$	0.1	\\
Tol~65			&	0.08247	$\pm$	0.00520	&	0.0011	$\pm$	0.0011	&	0.2503	$\pm$	0.0119	&	3.5	$\pm$	0.1	\\
SBS~1415+437~(No.~2) 	&	0.08610	$\pm$	0.00625	&	0.0000			&	0.2560	$\pm$	0.0138	&	4.2	$\pm$	0.1	\\
HS~1442+4250		&	0.08295	$\pm$	0.00676	&	0.0026	$\pm$	0.0026	&	0.2547	$\pm$	0.0161	&	4.4	$\pm$	0.2	\\
Mrk 209 		&	0.08475	$\pm$	0.00518	&	0.0010	$\pm$	0.0010	&	0.2552	$\pm$	0.0117	&	6.1	$\pm$	0.1	\\
Mrk~71~(No.~1)		&	0.08713	$\pm$	0.00580	&	0.0008	$\pm$	0.0008	&	0.2598	$\pm$	0.0128	&	7.2	$\pm$	0.2	\\
SBS~0917+527 		&	0.08664	$\pm$	0.00686	&	0.0020	$\pm$	0.0020	&	0.2613	$\pm$	0.0155	&	7.4	$\pm$	0.3	\\
SBS~1152+579		&	0.08906	$\pm$	0.00659	&	0.0012	$\pm$	0.0012	&	0.2648	$\pm$	0.0144	&	7.7	$\pm$	0.2	\\
SBS~1054+365 		&	0.09214	$\pm$	0.00850	&	0.0000			&	0.2688	$\pm$	0.0181	&	9.6	$\pm$	0.4	\\
SBS~0926+606 		&	0.09660	$\pm$	0.00987	&	0.0008	$\pm$	0.0008	&	0.2799	$\pm$	0.0205	&	10.0	$\pm$	0.4	\\
SBS~1135+581		&	0.08538	$\pm$	0.00382	&	0.0008	$\pm$	0.0008	&	0.2558	$\pm$	0.0086	&	11.7	$\pm$	0.3	\\
HS~0924+3821 		&	0.08509	$\pm$	0.00516	&	0.0000			&	0.2533	$\pm$	0.0115	&	12.3	$\pm$	0.5	\\
UM~439			&	0.09064	$\pm$	0.01247	&	0.0000			&	0.2654	$\pm$	0.0268	&	12.3	$\pm$	0.3	\\
%UM~396			&	0.08343	$\pm$	0.00530	&	0.0017	$\pm$	0.0017	&	0.2532	$\pm$	0.0124	&	18.1	$\pm$	0.7	\\
	
%SBS~0335-052E1		&	0.08629	$\pm$	0.00808	&	0.0023	$\pm$	0.0023	&	0.2616	$\pm$	0.0183	&	2.0	$\pm$	0.1	\\																
%HS~0122+0743		&	0.09960	$\pm$	0.00921	&	0.0007	$\pm$	0.0007	&	0.2861	$\pm$	0.0188	&	4.0	$\pm$	0.1	\\
%UM 461			&	0.09317	$\pm$	0.01198	&	0.0000			&	0.2712	$\pm$	0.0254	&	5.9	$\pm$	0.3	\\
%HS~0811+4913		&	0.10038	$\pm$	0.01168	&	0.0003	$\pm$	0.0003	&	0.2866	$\pm$	0.0237	&	9.4	$\pm$	0.2	\\
%%UM 238			&	0.08557	$\pm$	0.00920	&	0.0000			&	0.2542	$\pm$	0.0204	&	15.1	$\pm$	0.5	\\
																		
%UGC~4483		&	0.09639	$\pm$	0.00530	&	0.0000			&	0.2781	$\pm$	0.0110	&	3.5	$\pm$	0.1	\\
%SBS~1331+493		&	0.09058	$\pm$	0.00651	&	0.0000			&	0.2656	$\pm$	0.0140	&	5.8	$\pm$	0.2	\\
%SBS~1533+574B		&	0.09574	$\pm$	0.02033	&	0.0009	$\pm$	0.0009	&	0.2780	$\pm$	0.0422	&	12.6	$\pm$	0.6	\\
\hline
\end{tabular}
\caption{Primordial Helium Regression Values}
\label{table:PH}
\end{table}

\begin{table}[ht!]
\centering
\vskip .1in
\begin{tabular}{lcccc}
%\tabletypesize{\footnotesize}
%\tablewidth{0pt}
\hline\hline
Dataset							& N    	 & Y$_p$	 	& $dY/d(O/H)$	& $<Y>$ \\
\hline
Qualifying						& 14 	& 0.2534 $\pm$ 0.0083 	& 54 $\pm$ 102 	& 0.2574 $\pm$ 0.0036 \\
Qualifying + Flagged, excluding $\xi>0.01$		& 19	& 0.2580 $\pm$ 0.0074	& 16 $\pm$ 90	& 0.2592 $\pm$ 0.0034 \\
Qualifying + $\xi>0.01$ Flagged only			& 17 	& 0.2598 $\pm$ 0.0074	&  0 $\pm$ 95	& 0.2598 $\pm$ 0.0032 \\
Qualifying + All Flagged				& 22	& 0.2611 $\pm$ 0.0067	&  0 $\pm$ 86	& 0.2611 $\pm$ 0.0031 \\
Qualifying, $O/H < 9.2 \times 10^{-5}$ (AOS/AOS2)	&  9	& 0.2465 $\pm$ 0.0134	& 196 $\pm$ 230	& 0.2564 $\pm$ 0.0045 \\
\hline
\end{tabular}
\caption{Comparison of Y$_p$ for selected datasets}
\label{table:Yp's}
\end{table}

\section{Summary} \label{Conclusion}

Because of its importance as a probe of big bang cosmology, the predictions of BBN are under
constant scrutiny from new observations.  With the baryon-to-photon ratio fixed
by the WMAP determination of the baryon density, BBN leads to a distinct set of predictions for the
light element abundances. The prediction of the primordial D/H abundance is a major success when compared to determinations of D/H from quasar absorption system observations.
To further test the theory, accurate \he4 abundances are necessary as the BBN prediction for
\he4 is at the 0.1\% level.
Selecting an improved dataset from existing observations was a primary aim of this work. 

Unfortunately, \he4 abundance determinations from emission line fluxes carry 
significant systematic uncertainties.  In AOS2, we showed that one's ability to accurately estimate the set of physical parameters which are used to determine the \he4 abundance is greatly improved
with MCMC methods.  An eight-dimensional parameter space (which includes the helium abundance)
is sampled to calculate a set of line fluxes which can be compared directly with observed
fluxes. 

Here, we have applied the MCMC method to 
a large sample of 93 objects, and then carefully screened these objects to remove systematic bias, poor quality model fits, and spurious physical results.  In this process, the importance of He~I $\lambda$4026 was once again demonstrated.  Data lacking this line was found to be susceptible to a systematic increase in underlying Helium absorption and a corresponding bias to larger abundances. 
70 of 93 objects in the initial sample contain the $\lambda$4026 line, and were the subject of our 
further analysis.

One of the benefits of the MCMC method is our ability to directly test the goodness-of-fit
for a particular set of solutions.  While we are able to find a best fit solution almost every object,
there is no guarantee that that solution is actually a good (in a statistical sense) description of the data.
As the best fit solution (fitting nine observables with eight parameters) is found by minimizing 
$\chi^2$, the resulting minimum $\chi^2$ is a direct measure of the goodness-of-fit. 
Unfortunately, 45 of the 70 objects resulted in a $\chi^2$ per degree of freedom $> 4$. 
In our quest for high quality and reliable \he4 data, these data were excluded from further analysis.
On a positive note, the practical impact of the 95\% confidence level cut preferentially removed outlier parameter solutions. Indeed, only a handful of objects surviving the $\chi^2$ were flagged as outliers.  

The cumulative effect of the careful selection led to a dataset with 14 objects.  However, the reliability of this Final Dataset has been methodically evaluated and is a distinct improvement over previous efforts. The precision of the resulting primordial helium abundance is increased, even if not to the level we hope to someday achieve.  The concordance of the various regression datasets incorporating flagged points underscores the robustness of the result and the improved quality of the dataset.  

As we have emphasized before, the most promising avenue for significant future improvement in the primordial helium abundance determination lies with higher quality, high resolution spectra.  Decreased measurement uncertainties will clearly better constrain the solution and decrease systematic uncertainty.  Beyond that straightforward benefit, there are a couple of promising, related paths.  First, as discussed in AOS, high resolution spectra afford the chance to measure the absorption underlying the Balmer lines directly.  Removing underlying hydrogen absorption as a solution model parameter would reduce degeneracies and produce better constrained parameters.  Similarly, high signal to noise spectra open up the possibility of adding a weaker Helium or Hydrogen line to our analysis.  Again, increasing the number of lines constraining the parameters would reduce degeneracies and decrease systematic uncertainty.  That both of these improvements would directly benefit the neutral hydrogen fraction, the least well constrained parameter, provides additional impetus.  Finally, we also plan to quantify the correlations between the parameters for each object.  This will yield further insight into the goodness-of-fit.

In summary, we have demonstrated the rigor and transparency of MCMC methods in selecting the best available data to ultimately extract the primordial He abundance.  It supports a stringent screening of candidate spectra, yielding a robust sample.  The fruit of these labors is an improved determination of the primordial helium abundance, both in increased confidence in its accuracy and in a modest increase in its precision.  These uncertainties are still relatively large, however, and the case for needing higher quality spectra is further strengthened.

\acknowledgments

We wish to thank Yuri Izotov for encouraging us to expand our analysis to a larger dataset and
for providing the $\lambda$4026 measurements for the HeBCD sample.
The work of EA and KAO is supported in part by DOE grant DE-FG02-94ER-40823. 
EDS is grateful for partial support from the University of Minnesota.


\newpage
\begin{thebibliography}{}


\bibitem[Walker et al.(1991)]{wssok} 
  T.~P.~Walker, G.~Steigman, D.~N.~Schramm, K.~A.~Olive and H.~S.~Kang,
  %``Primordial nucleosynthesis redux,''
  Astrophys.\ J.\  {\bf 376}, 51 (1991).
  %%CITATION = ASJOA,376,51;%%

\bibitem[Olive, Steigman, \& Walker(2000)]{osw}
  K.~A.~Olive, G.~Steigman and T.~P.~Walker,
  %``Primordial Nucleosynthesis: Theory and Observations,''
  Phys.\ Rept.\  {\bf 333}, 389 (2000)
  [arXiv:astro-ph/9905320].
  %%CITATION = PRPLC,333,389;%%

\bibitem[Fields \& Sarkar(2008)]{fs} 
  B.~D.~Fields and S.~Sarkar,
  %``Big-Bang nucleosynthesis,''
  in K. Nakamura et al.
  J.\ Phys.\ G {\bf 37}, 075021 (2010)
%\href{http://www.slac.stanford.edu/spires/find/hep/www?irn=8096686}{SPIRES entry}


\bibitem[Komatsu et al.(2009)]{wmap} 
  E.~Komatsu {\it et al.}  [WMAP Collaboration],
  %``Five-Year Wilkinson Microwave Anisotropy Probe (WMAP\altaffilmark 1 )
  %Observations:Cosmological Interpretation,''
  Astrophys.\ J.\ Suppl.\  {\bf 180}, 330 (2009)
  [arXiv:0803.0547 [astro-ph]].
  %%CITATION = APJSA,180,330;%%

\bibitem[Komatsu et al.(2010)]{wmap10} 
   E.~Komatsu {\it et al.} [WMAP Collaboration],
  %``Seven-Year Wilkinson Microwave Anisotropy Probe (WMAP) Observations: Cosmological Interpretation,''
  Astrophys.\ J.\ Suppl.\ \ {\bf 192}, 18  (2011)
  [arXiv:1001.4538 [astro-ph.CO]].
  %%CITATION = APJSA,192,18;%%
    
  \bibitem[Cyburt et al.(2002)]{cfo2}
  R.~H.~Cyburt, B.~D.~Fields and K.~A.~Olive,
  %``Primordial nucleosynthesis with CMB inputs: Probing the early universe and light element astrophysics,''
  Astropart.\ Phys.\ \ {\bf 17}, 87  (2002)
  [astro-ph/0105397].
  %%CITATION = APHYE,17,87;%%

\bibitem[Cyburt et al.(2001)]{cfo} 
  R.~H.~Cyburt, B.~D.~Fields and K.~A.~Olive,
  %``The NACRE Thermonuclear Reaction Compilation and Big Bang
  %Nucleosynthesis,''
  New Astron.\  {\bf 6}, 215 (2001)
  [arXiv:astro-ph/0102179].
  %%CITATION = NEWAS,6,215;%%

\bibitem[Coc et al.(2002)]{coc} 
  A.~Coc, E.~Vangioni-Flam, M.~Casse and M.~Rabiet,
  %``Constraints on Omega b from nucleosynthesis of Li-7 in the standard big
  %bang model,''
  Phys.\ Rev.\  D {\bf 65}, 043510 (2002)
  [arXiv:astro-ph/0111077].
  %%CITATION = PHRVA,D65,043510;%%
  
  \bibitem[Cyburt et al.(2003)]{cfo3}
  R.~H.~Cyburt, B.~D.~Fields and K.~A.~Olive,
  %``Primordial nucleosynthesis in light of WMAP,''
  Phys.\ Lett.\ B\ {\bf 567}, 227  (2003)
  [astro-ph/0302431].
  %%CITATION = PHLTA,B567,227;%%

\bibitem[Coc et al.(2004)]{coc2} 
  A.~Coc, E.~Vangioni-Flam, P.~Descouvemont, A.~Adahchour and C.~Angulo,
  %``Updated Big Bang Nucleosynthesis confronted to WMAP observations and to the
  %Abundance of Light Elements,''
  Astrophys.\ J.\  {\bf 600}, 544 (2004)
  [arXiv:astro-ph/0309480].
  %%CITATION = ASJOA,600,544;%%

\bibitem[Cyburt(2004)]{cyburt} 
  R.~H.~Cyburt,
  %``Primordial Nucleosynthesis for the New Cosmology: Determining Uncertainties
  %and Examining Concordance,''
  Phys.\ Rev.\  D {\bf 70}, 023505 (2004)
  [arXiv:astro-ph/0401091].
  %%CITATION = PHRVA,D70,023505;%%

\bibitem[Descouvemont et al.(2004)]{coc3} 
  P.~Descouvemont, A.~Adahchour, C.~Angulo, A.~Coc and E.~Vangioni-Flam,
  %``Compilation and R-matrix analysis of big bang nuclear reaction rates,''
  Atomic Data and Nuclear Data Tables, {\bf 88}, 203 (2004)
  [arXiv:astro-ph/0407101].
  %%CITATION = ASTRO-PH/0407101;%%

\bibitem[Cuoco et al.(2004)]{cuoco} 
  A.~Cuoco, F.~Iocco, G.~Mangano, G.~Miele, O.~Pisanti and P.~D.~Serpico,
  %``Present status of primordial nucleosynthesis after WMAP: results from a new
  %BBN code,''
  Int.\ J.\ Mod.\ Phys.\  A {\bf 19}, 4431 (2004)
  [arXiv:astro-ph/0307213].
  %%CITATION = IMPAE,A19,4431;%%

\bibitem[Serpico et al.(2004)]{serp} 
  P.~D.~Serpico, S.~Esposito, F.~Iocco, G.~Mangano, G.~Miele and O.~Pisanti,
  %``Nuclear Reaction Network for Primordial Nucleosynthesis: a detailed
  %analysis of rates, uncertainties and light nuclei yields,''
  JCAP {\bf 0412}, 010 (2004)
  [arXiv:astro-ph/0408076].
  %%CITATION = JCAPA,0412,010;%%

\bibitem[Cyburt et al.(2008)]{cfo5} 
  R.~H.~Cyburt, B.~D.~Fields and K.~A.~Olive,
  %``A Bitter Pill: The Primordial Lithium Problem Worsens,''
  JCAP {\bf 0811}, 012 (2008)
  [arXiv:0808.2818 [astro-ph]].
  %%CITATION = JCAPA,0811,012;%%

\bibitem[Coc et al.(2011)]{coc4} 
  A.~Coc, S.~Goriely, Y.~Xu, M.~Saimpert and E.~Vangioni,
  %``Standard Big-Bang Nucleosynthesis up to CNO with an improved extended
  %nuclear network,''
  arXiv:1107.1117 [astro-ph.CO].
  %%CITATION = ARXIV:1107.1117;%%

\bibitem[Nakamura et al.(2010)]{rpp}  
  K.~Nakamura {\it et al.}  [Particle Data Group],
  %``Review of particle physics,''
  J.\ Phys.\ G {\bf 37}, 075021 (2010).
  %%CITATION = JPHGB,G37,075021;%%

\bibitem[Peimbert \& Torres-Peimbert(1974)]{ptp74} 
  M.~Peimbert and S.~Torres-Peimbert,
  Astrophys.\ J.\  {\bf 193}, 327 (1974)

\bibitem[Olive \& Skillman(2001)]{os01}  
  K.~A.~Olive and E.~D.~Skillman,
  %``On the Determination of the He Abundance in Extragalactic H II Regions,''
  New Astron.\ {\bf 6}, 119 (2001)
  arXiv:astro-ph/0007081.
  %%CITATION = ASTRO-PH/0007081;%%

\bibitem[Olive \& Skillman(2004)]{os04}  
  K.~A.~Olive and E.~D.~Skillman,
  %``A Realistic Determination of the Error on the Primordial Helium Abundance:
  %Steps Toward Non-Parametric Nebular Helium Abundances,''
  Astrophys.\ J.\  {\bf 617}, 29 (2004)
  [arXiv:astro-ph/0405588].
  %%CITATION = ASJOA,617,29;%%

\bibitem[Izotov, Thuan, \& Stasi\'nska(2007)]{its07} 
  Y.~I.~Izotov, T.~X.~Thuan and G.~Stasi\'nska,
  %``The primordial abundance of 4He: a self-consistent empirical analysis of
  %systematic effects in a large sample of low-metallicity HII regions,''
  Astrophys.\ J.\  {\bf 662}, 15 (2007)
  [arXiv:astro-ph/0702072].
  %%CITATION = ASJOA,662,15;%%

\bibitem[Aver et al.(2011)]{AOS2}
  E.~Aver, K.~A.~Olive and E.~D.~Skillman,
  %``Mapping systematic errors in helium abundance determinations using Markov
  %Chain Monte Carlo,''
  JCAP {\bf 1103}, 043 (2011)
  [arXiv:1012.2385 [astro-ph.CO]] (AOS2).
  %%CITATION = JCAPA,1103,043;%%

\bibitem[Izotov, Thuan, \& Lipovetsky(1994)]{itl94}
  Y.~I.~Izotov, T.~X.~Thuan and V.~A.~Lipovetsky,
  %``The Primordial helium abundance from a new sample of metal-deficient blue
  %compact galaxies,''
  Astrophys.\ J.\  {\bf 435}, 647 (1994).
  %%CITATION = ASJOA,435,647;%%

\bibitem[Peimbert, Peimbert, \& Ruiz(2000)]{ppr00}
  M.~Peimbert, A.~Peimbert and M.~T.~Ruiz,
  %``The Chemical Composition of the Small Magellanic Cloud H II Region NGC 346
  %and the Primordial Helium Abundance,''
  Astrophys.\ J.\  {\bf 541}, 688 (2000)
  [arXiv:astro-ph/0003154].
  %%CITATION = ASTRO-PH/0003154;%%

\bibitem[Aver et al.(2010)]{AOS} 
  E.~Aver, K.~A.~Olive and E.~D.~Skillman,
  %``A New Approach to Systematic Uncertainties and Self-Consistency in Helium
  %Abundance Determinations,''
  JCAP {\bf 1005}, 003 (2010)
  [arXiv:1001.5218 [astro-ph.CO]] (AOS).
  %%CITATION = JCAPA,1005,003;%%

\bibitem[Izotov \& Thuan(2004)]{it04}
  Y.~I.~Izotov and T.~X.~Thuan,
  %``Systematic effects and a new determination of the primordial abundance of
  %He-4 and dY/dZ from observations of blue compact galaxies,''
  Astrophys.\ J.\  {\bf 602}, 200 (2004)
  [arXiv:astro-ph/0310421].
  %%CITATION = ASJOA,602,200;%%

%\cite{Asplund:2009fu}
\bibitem[Asplund et al.(2009)]{asp09}
  M.~Asplund, N.~Grevesse, A.~J.~Sauval and P.~Scott,
  %``The chemical composition of the Sun,''
  Ann.\ Rev.\ Astron.\ Astrophys.\  {\bf 47}, 481 (2009)
  [arXiv:0909.0948 [astro-ph.SR]].
  %%CITATION = ARAAA,47,481;%%

\bibitem[Peimbert et al.(2007)]{plp07} 
  M.~Peimbert, V.~Luridiana, and A.~Peimbert,
  %``Revised Primordial Helium Abundance Based on New Atomic Data,''
  Astrophys.\ J.\ {\bf 666}, 636  (2007)
  [arXiv:astro-ph/0701580].

\bibitem[Benjamin, Skillman, \& Smits(2002)]{bss02}
  R.~A.~Benjamin, E.~D.~Skillman and D.~P.~Smits,
  %``Radiative Transfer Effects in He I Emission Lines,''
  Astrophys.\ J.\  {\bf 569}, 288 (2002)
  [arXiv:astro-ph/0202227].
  %%CITATION = ASTRO-PH/0202227;%%


\end{thebibliography}
\end{document}